%% file: ndss2021.tex
\documentclass[conference]{IEEEtran}

\usepackage[table,usenames,dvipsnames]{xcolor} % More colors
\usepackage[nolist]{acronym} % acronym package
\usepackage[hyphens]{url} % break very long urls not only by / but also by -
\usepackage{hyperref} % url
\usepackage{graphicx} % For figures (resize)
\usepackage{subcaption} % Syntax between subfigure (old) and subcaption (new) is different
\usepackage{booktabs} % For formal tables
\usepackage{multirow} % Multirows in tables
\usepackage{CJKutf8}  % Required for Chinese symbols (wake words)
\usepackage{paralist} % For compact itemize and enumerate
\usepackage{balance}  % Balance references
\usepackage[pscoord]{eso-pic} % Alexa stop at the end of the paper
\usepackage{amsmath}
\usepackage{cite}
\usepackage{float} % ULTRA EVIL

\usepackage{xspace} % For i.e., e.g., macros
\newcommand{\ie}{i.\@\,e.,\@\xspace}
\newcommand{\eg}{e.\@\,g.,\@\xspace}

\newcommand{\etal}{et~al.\@\xspace}

\makeatletter
\newcommand{\paragraphtiny}{%
  \@startsection{paragraph}{4}%
  {\z@}{1.00ex \@plus 1ex \@minus .2ex}{-1em}%
  {\normalfont\normalsize\bfseries}%
}
\makeatother

\makeatletter
\renewcommand{\paragraph}{%
  \@startsection{paragraph}{4}%
  {\z@}{1.00ex \@plus 1ex \@minus .2ex}{-1em}%
  {\normalfont\normalsize\bfseries}%
}
\makeatother

\begin{document}

\input{acronyms}

\title{\emph{``Unacceptable, where is my privacy?''}\\Exploring Accidental Triggers of Smart Speakers}

\author{
\IEEEauthorblockN{Lea Sch\"onherr\authorrefmark{1}, Maximilian Golla\authorrefmark{2}, Thorsten Eisenhofer\authorrefmark{1}, Jan Wiele\authorrefmark{1}, Dorothea Kolossa\authorrefmark{1}, and Thorsten Holz\authorrefmark{1}}
\IEEEauthorblockA{\authorrefmark{1}Ruhr University Bochum; \authorrefmark{2}Max Planck Institute for Security and Privacy\\
\{lea.schoenherr,thorsten.eisenhofer,jan.wiele,dorothea.kolossa,thorsten.holz\}@rub.de; maximilian.golla@csp.mpg.de}
}

\maketitle

\begin{abstract}
Voice assistants like Amazon's Alexa, Google's Assistant, or Apple's Siri, have become the primary (voice) interface in smart speakers that can be found in millions of households. For privacy reasons, these speakers analyze every sound in their environment for their respective \emph{wake word} like ``Alexa'' or ``Hey Siri,'' before uploading the audio stream to the cloud for further processing. Previous work reported on the inaccurate wake word detection, which can be tricked using similar words or sounds like ``cocaine noodles'' instead of ``OK Google.''

In this paper, we perform a comprehensive analysis of such \emph{accidental triggers}, i.\,e., sounds that should not have triggered the voice assistant, but did. More specifically, we automate the process of finding accidental triggers and measure their prevalence across 11~smart speakers from 8 different manufacturers using everyday media such as TV shows, news, and other kinds of audio datasets. To systematically detect accidental triggers, we describe a method to artificially craft such triggers using a pronouncing dictionary and a weighted, phone-based Levenshtein distance. In total, we have found hundreds of accidental triggers. Moreover, we explore potential gender and language biases and analyze the reproducibility. Finally, we discuss the resulting privacy implications of accidental triggers and explore countermeasures to reduce and limit their impact on users' privacy. To foster additional research on these sounds that mislead machine learning models, we publish a dataset of more than 1000 verified triggers as a research artifact.
\end{abstract}

\input{introduction.tex}

\input{misc.tex}
\input{setup.tex}
\input{prevalence.tex}
\input{crafting.tex}
\input{related.tex}
\input{discussion.tex}
\input{conclusion.tex}

\clearpage
\balance
\bibliographystyle{IEEEtranS} 
\bibliography{references}

\end{document}

%% file: acronyms.tex
\begin{acronym}
    \acro{TTS}{text-to-speech}
    \acro{WSJ}{\emph{WallStreetJournal}}
    \acro{ASR}{automatic speech recognition}
\end{acronym}

%% file: introduction.tex
\section{Introduction}
In the past few years, we have observed a huge growth in the popularity of voice assistants, especially in the form of smart speakers. All major technology companies, among them Amazon, Baidu, Google, Apple, and Xiaomi, have developed an assistant. Amazon is among the most popular brands on the market: the company reported in 2019 that it had sold more than $100$~million devices with \emph{Alexa} on board; there were more than $150$~products that support this voice assistant (e.\,g., smart speakers, soundbars, headphones, etc.) ~\cite{bohn-19-100m-alexa-devices}. Especially smart speakers are on their way of becoming a pervasive technology, with several security and privacy implications due to the way these devices operate: they continuously analyze every sound in their environment in an attempt to recognize a so-called \emph{wake word} such as ``Alexa,'' ``Echo,'' ``Hey Siri,'' or ``Xiǎo dù xiǎo dù.'' Only if a wake word is detected, the device starts to record the sound and uploads it to a remote server, where it is transcribed, and the detected word sequence is interpreted as a command. This mode of operation is mainly used due to privacy concerns, as the recording of all (potentially private) communication and processing this data in the cloud would be too invasive. Furthermore, the limited computing power and storage on the speaker prohibits a full analysis on the device itself. Hence, the recorded sound is sent to the cloud for analysis once a wake word is detected.

Unfortunately, the precise sound detection of wake words is a challenging task with a typical trade-off between usability and security: manufacturers aim for a low false acceptance and false rejection rate~\cite{raju-18-robust-wake-word}, which enables a certain wiggle room for an adversary. As a result, it happens that these smart speaker trigger even if the wake word has not been uttered. First explorative work on the confusion of voice-driven user input has been done by Vaidya et al.~\cite{vaidya-15-cocaine}. In their 2015 paper, the authors explain how Google's voice assistant, running on a smartphone \emph{misinterprets} ``cocaine noodles'' as ``OK Google'' and describe a way to exploit this behavior to execute unauthorized commands such as sending a text, calling a number, or opening a website.
Later, Kumar et al.~\cite{kumar-18-skillsquatting} presented an attack, called \emph{skill squatting}, that leverages transcription errors of a list of similar-sounding words to existing Alexa skills. Their attack exploits the \emph{imperfect transcription} of the words by the Amazon API and routes users to malicious skills with similar-sounding names. A similar attack, in which the adversary exploits the way a skill is invoked, has been described by Zhang et al.~\cite{zhang-19-dangerousskills}. 

Such research results utilize instances of what we call an \emph{accidental trigger}: a sound that a voice assistant mistakes for its wake word. Privacy-wise, this can be fatal, as it will induce the voice assistant to start a recording and stream it to the cloud. Inadvertent triggering of smart speakers and the resulting accidentally captured conversations are seen by many as a privacy threat~\cite{malkin-19-priv-attitudes,chung-17-alexa-trust,dubois-20-characterizing-misactivations}. When the media reported in summer 2019 that employees of the manufacturer listen to voice recordings to transcribe and annotate them, this led to an uproar~\cite{day-19-amazon-workers-listening,van-hee-19-google-employees-eavesdropping}. As a result, many companies paused these programs and no longer manually analyze the recordings~\cite{lecher-19-google-pause-listening,hern-19-amazon-disable-human-review,gartenberg-19-apple-apologizes}.

In this paper, we perform a systematic and comprehensive analysis of accidental triggers to understand and elucidate this phenomenon in detail. To this end,  we propose and implement an automated approach for systematically evaluating the resistance of smart speakers to such accidental triggers. 
We base this evaluation on candidate triggers carefully crafted from a pronouncing dictionary with a novel phonetic distance measure, as well as on available AV media content and bring it to bear on a range of current smart speakers. More specifically, in a first step, we analyze vendor's protection mechanisms such as cloud-based wake word verification systems and acoustic fingerprints, used to limit the impact of accidental triggers. We carefully evaluate how a diverse set of $11$~smart speakers from $8$~manufacturers behaves in a simulated living-room-like scenario with different sound sources (\eg TV shows, news, and professional audio datasets). We explore the feasibility of artificially crafting accidental triggers using a pronouncing dictionary and a weighted, phone-based Levenshtein distance metric and benchmark the robustness of the smart speakers against such crafted accidental triggers. We found that a distance measure that considers phone-dependent weights is more successful in describing potential accidental triggers. Based on this measure, we crafted 1-, 2-, and 3-grams as potential accidental triggers, using a \ac{TTS} service and were able to find accidental triggers for all tested smart speakers in a fully automated way.

Finally, we give recommendations and discuss countermeasures to reduce the number of accidental triggers or limit their impact on users' privacy.

% CONTRIBUTIONS
\smallskip
\noindent
To summarize, we make the following key contributions:

\begin{enumerate}%[noitemsep,nolistsep]
    \item By reverse-engineering the communication channel of an Amazon Echo, we are able to provide novel insights on how commercial companies deal with such problematic triggers in practice.
    \item We develop a fully automated measurement setup that enables us to perform an extensive study of the prevalence of accidental triggers for $11$~smart speakers from $8$~manufacturers. We analyze a diverse set of audio sources, explore potential gender and language biases, and analyze the identified triggers' reproducibility.
    \item We introduce a method to synthesize accidental triggers with the help of a pronouncing dictionary and a weighted phone-based Levenshtein distance metric. We demonstrate that this method enables us to find new accidental triggers in a systematic way and argue that this method can benchmark the robustness of smart speakers.
    \item We publish a dataset of more than 1000 accidental triggers to foster future research on this topic.\footnote{They are available at: \textcolor{NavyBlue}{ \url{https://unacceptable-privacy.github.io}}}
\end{enumerate}

%% file: misc.tex
\section{Understanding Accidental Triggers}\label{sec:understanding-acc-triggers}

In this section, we provide the required background on wake word detection. Furthermore, we describe how Amazon deals with accidental triggers and how we analyzed and reverse engineered an Amazon Echo speaker and end with an overview of smart speaker privacy settings.

\subsection{Wake Word Recognition}
To enable natural communication between the user and the device, automatic speech recognition (ASR) systems built into smart speakers rely on a far-field voice-based activation. In contrast to a push-to-talk model, where speech recognition is only active after a physical button is pressed, smart speakers continuously record their surroundings to allow hands-free use. After detecting a specific \emph{wake word}, also known as hotword or keyword, the smart speaker starts to respond.
The wake word recognition system is often a lightweight DNN-based ASR system, limited to a few designated words~\cite{guo-18-echo-wake-word,wu-18-echo-wake-word,apple-17-siri-cbwwv}. To guarantee the responsiveness, the recognition runs locally and is limited by the computational power and storage of the speaker. For example, we found the \emph{Computer} wake word model for an Amazon Echo speaker to be less than $2$~MB in size, running on a 1GHz ARM Cortex-A8 processor. The speaker uses about 50\% of its CPU time for the wake word recognition process. In addition to the wake word, the model also detects a stop signal (``Stop'') to interrupt the currently running request. Especially when used in environments with ambient noise from external sources such as TVs, a low false acceptance and false rejection rate is most important for these systems~\cite{raju-18-robust-wake-word}.

The device will only transmit data to the respective server \emph{after} the wake word has been recognized. Hence, activating the wake word by an accidental trigger will lead to the upload of potentially sensitive and private audio data, and should, therefore, be avoided as far as possible.

In some cases, a speaker misinterprets another word or sound as its wake word. If the misinterpreted word is unrelated to the configured wake word, we refer to this event as an \emph{accidental trigger}. To limit the consequences of such false wakes, vendors started to augment the local wake word recognition with a \emph{cloud-based wake word verification}. Moreover, there is an \emph{acoustic fingerprint}-based mechanism in place that prevents a speaker from triggering when listening to certain audio sequences observed in TV commercials and similar audio sources. We describe both of these mechanisms in more detail in Section~\ref{sec:dealing-with-triggers}.

\subsection{Voice Profiles and Sensitivity}\label{sec:voice-profiles}
Voice profiles, also referred to as ``Voice Match'' or ``Recognize My Voice'' feature, are a convenience component of modern voice assistants~\cite{welch-17-alexa-voice-profiles}. The requisite voice training was introduced with iOS 9 (2015), and Android 8 (2017) to build context around questions and deliver personalized results. On smartphones, a voice profile helps to recognize the user better~\cite{apple-18-personalized-siri}. Vendors explain that without a profile, queries are simply considered to be coming from guests and thus will not include personal results~\cite{google-20-voice-match-personal-results}.

In contrast to voice assistants on phones, smart speakers are intended to be activated by third parties, such as friends and visitors. Thus, voice profiles do not influence whether a smart speaker is activated or not when the wake word is recognized. In shared, multi-user environments, voice profiles enable voice assistants to tell users apart and deliver personalized search results, music playlists, and communication. The feature is also not meant for security as a similar voice or recording can trick the system~\cite{gebhart-17-fooling-voice-match}. In our experiments, voice profiles were not enabled or used.

In April 2020, Google introduced a new feature that allows users to adjust the wake word's responsiveness to limit the number of accidental activations~\cite{google-20-sensitivity}. In our experiments, we used the ``Default'' sensitivity.

\subsection{Alexa Internals}
In the following, we describe how we analyzed and reverse engineered an Amazon Echo speaker (1st Gen.). The speaker was bought in February 2017 and was equipped with firmware version 647 588 720 from October 2019.

\paragraph{Rooting} % an Amazon Echo}
To obtain root access on an Amazon Echo speaker, we follow a method described by Clinton et al.~\cite{clinton-16-analyzing-echo} that was later refined by Barnes~\cite{barnes-17-fsecure-alexa-listening}. To decrypt and analyze the speaker's communication with the Alexa API, we inject a shared object that dumps all negotiated pre-master secrets into a file, which we later use to decrypt the TLS protected traffic recorded in the PCAP files using Wireshark.

\paragraph{From the Wake Word to the Response}\label{sec:root-alexa:wakeword-to-response}
The communication between the Echo smart speaker and Amazon's cloud is handled by the \emph{AlexaDaemon} using the latency-optimized SPDY protocol. The first step is to authenticate the device. After a successful authentication, the cloud requests various information, e.\,g., volume levels, information about currently playing songs, a list of paired Bluetooth devices, the used Wi-Fi network, and the configured wake word, language, and~timezone.

Echo's ASR engine is called \emph{Pryon} and started from a fork of the open-source speech recognition toolkit Kaldi~\cite{povey-11-kaldi}. On our Amazon Echo speaker, we found four wake words models \emph{Alexa}, \emph{Computer}, \emph{Echo}, and \emph{Amazon}, divided by the two different device types \emph{doppler} and \emph{pancake} (Echo and Echo~Dot), and four different languages/regions (en-US, es-US, en-GB, and de-DE). The local automatic speech recognition daemon, ASRD, uses Pryon to detect the configured wake word. The ASRD represents its certainty for recognizing the wake word with a \emph{classifier score} between 0.0 and 1.0.

The following six aspects are relevant for us: \raisebox{.5pt}{\textcircled{\raisebox{-.9pt} {1}}} In normal use, a wake word score above $0.57$ is categorized as an ``Accept.'' A score between $0.1$ (notification threshold) and $0.57$ will be categorized as a ``NearMiss.'' The classifier threshold for an accept is lowered to $0.43$, if the device is playing music or is in a call. A near miss will not trigger the LED indicator or any components, and no audio will be processed or uploaded to the Amazon cloud.

\raisebox{.5pt}{\textcircled{\raisebox{-.9pt} {2}}} In contrast, an ``Accept'' will activate the AudioEncoderDaemon, which encodes the currently recorded audio. At the same time, the Echo speaker switches to the \emph{SendingDataToAlexa} state. The LED indicator turns on and starts to indicate the estimated direction of the speech source. Moreover, the AlexaDaemon informs the cloud about an upcoming audio stream, together with the information where in the stream the ASRD believes to have recognized the wake word.

\raisebox{.5pt}{\textcircled{\raisebox{-.9pt} {3}}} The Amazon cloud then runs its own detection of the wake word (cf. Section~\ref{sec:cbwwv}).
\raisebox{.5pt}{\textcircled{\raisebox{-.9pt} {4}}} If the cloud recognizes the wake word, it will send a transcription of the identified question, e.\,g., ``what are the Simpsons.'' In response, Echo will switch to the \emph{AlexaThinking} mode and the LED indicator will change to a blue circulating animation. In the meantime, the conversational intelligence engine in the cloud tries to answer the question. 

\raisebox{.5pt}{\textcircled{\raisebox{-.9pt} {5}}} Next, the Amazon cloud will respond with the spoken answer to the question, encoded as an MP3~file, played using the AlexaSpeechPlayer. Echo then notifies the cloud that it is playing, and the LED indicator switches to a blue fade in/out animation, while the speaker switches to the \emph{TTSFromAlexa} state. At the same time, the server requests the device to stop sending the microphone input.
\raisebox{.5pt}{\textcircled{\raisebox{-.9pt} {6}}} After the AlexaSpeechPlayer finished, the AlexaDaemon informs the cloud about the successful playback, and Echo switches the LED indicator off by changing to the \emph{AlexaDialogEnd} state.

Summarizing, we can confirm that the examined device is only transmitting microphone input to Amazon's cloud if the LED indicator is active and hence acting as a trustworthy indicator for the user. Based on a packet flow analysis, this is also true for all other voice assistants. One exception is the smart speaker built by Xiaomi, which seems to upload speech that can be considered a near miss to overrule the local ASR engine, without switching on the LED indicator.

\subsection{Reducing Accidental Triggers}\label{sec:dealing-with-triggers}
Next, we focus on two methods that vendors deployed in an attempt to prevent or recover from accidental triggers.

\subsubsection{Cloud-Based Wake Word Verification}\label{sec:cbwwv}
As described above, the local speech recognition engine is limited by the speaker's resources. Thus, in May 2017, Amazon deployed a two-stage system~\cite{karczewski-17-alexa-cbwwv}, where a low power ASR on the Echo is supported by a more powerful ASR engine in the cloud.

Accordingly, accidental triggers can be divided into two categories: (i) \emph{local triggers} that overcome the local classifier, but get rejected by the cloud-based ASR engine, and (ii) \emph{local + cloud triggers} that overcome both. While a local trigger switches the LED indicator on, a subsequent question ``\emph{\{accidental local trigger\}}, will it rain today?'' will not be answered.
In cases where the cloud does not confirm the wake word's presence, it sends a command to the Echo to stop the audio stream. Surprisingly, the entire process from the local recognition of the wake word to the moment where Echo stops the stream and switches off the LED indicator only takes about $1-2$~seconds. In our tests, we observe that during this process, Echo uploads at least $1-2$ seconds of voice data, approx. $0.5$~seconds of audio before the detected wake word occurs, plus the time required to utter the wake word (approx. another second). In cases where the cloud-based ASR system also detects the wake word's presence, the accidental trigger can easily result in the upload of $10$ or more seconds of voice data.
During our experiments, we found that all major smart speaker vendors use a cloud-based verification system, including Amazon, Apple, Google, and Microsoft.

\subsubsection{Acoustic Fingerprints}
To prevent TV commercials and similar audio sources from triggering Echo devices, Amazon uses an acoustic fingerprinting technique. Before the device starts to stream the microphone input to the cloud, a local database of fingerprints is evaluated. In the cloud, the audio is checked against a larger set of fingerprints. The size of the local database on an Amazon Echo (1st Gen.) speaker is limited by its CPU power and contains $50$~entries. This database gets updated approximately every week with 40 new fingerprints, which mostly contain currently airing advertisements~\cite{rodehorst-19-alexa-super-bowl-ad} for Amazon products. Until mid-May, the database contained dates and clear text descriptions of the entries. Since then, only hash values are stored. The database of Mai 2020 still contained 9~fingerprints from 2017, e.\,g., the ``Echo Show: Shopping Lists'' YouTube video. 

We evaluate some of the entries by searching for the commercials and find videos where the ASRD reports a ``[...] strong fingerprint match.'' However, we also observed false positives, where fingerprint matches against commercials were found, which were not present in the database, leaving the robustness of the technique~\cite{haitsma-02-acoustic-fingerprint} in question.
We found that the metricsCollector process on the Echo speaker periodically collects and uploads the detected fingerprints. This is particularly concerning for privacy since it shows an interest of Amazon in these local fingerprint matches that could easily be combined with the cloud matches and be abused to build viewing profiles of the users~\cite{mohajeri-19-tv-streaming-tracking}. If the wake word is spoken on live TV, Amazon will register a large peak in concurrent triggers with the same audio fingerprint and automatically request the devices to stop~\cite{rodehorst-19-alexa-super-bowl-ad}.

\subsection{Smart Speaker Privacy Settings}\label{sec:privacy-settings}
To learn more about how vendors handle their users' data, we requested the voice assistant interaction history from Amazon, Apple, Google, and Microsoft using their respective web forms. Among the tested vendors, Apple is the only manufacturer that does not provide access to the voice data but allows users to request their complete deletion.

\begin{table}[tb]
  \centering
  \caption{Smart Speaker Privacy Settings}
  \vspace{-1em}
  \smallskip
    \begin{tabular}{r|cllc|c}
    \toprule
    & \multicolumn{4}{c|}{\textbf{Voice Recordings}} & \textbf{Local} \\
    \textbf{Vendor} & \textbf{Opt-Out} & \textbf{Retention} & \textbf{Delete} & \textbf{Report} & \textbf{Trigger} \\
    \toprule
    Amazon    & Yes~      & 3, 18 months       & A, R, I & Yes & Yes \\
    Apple     & Yes~      & 6, 24 months       & A       & -   & - \\
    Google    & Yes~      & 3, 18 months       & A, R, I & No  & Yes \\
    Microsoft & Yes*      & Unspecified        & A, I    & No  & No \\
    \bottomrule
    \end{tabular}\\
    \smallskip
    {\footnotesize
    *Cannot speak to Cortana anymore; A=All, R=Range, I=Individual.
    }
    \label{tab:dashboard}
\end{table}

In Table~\ref{tab:dashboard}, we analyzed if the user is able to opt-out of the automatic storing of their voice data, how long the recordings will be retained, the possibility to request the deletion of the recordings, and whether recordings could be reported as problematic. Furthermore, we checked if false activations through accidental triggers, i.\,e., local triggers, are visible to the user (``Audio was not intended for Alexa'').
Apple reports to store the voice recordings using a device-generated random identifier for up to 24 months but promises to disassociate the recordings from related request data (location, contact details, and app data)~\cite{hern-19-siri-addressbook-apps-location}, after six months. In contrast, customers of Amazon and Google can choose between two different voice data retention options. According to Google, the two time frames of 3 and 18-months are owed to \emph{recency} and \emph{seasonality}~\cite{pierce-20-google-3-and-18-months}. Microsoft's retention policy is more vague, but they promise to comply with legal obligations and to only store the voice data ``as long as necessary.''

%% file: setup.tex
\section{Evaluation Setup}
In this section, we describe our evaluated smart speakers and the datasets we used for our measurement study.

\begin{table*}[t]
  \centering
  \caption{Evaluated Smart Speakers}
  \vspace{-1em}
  \smallskip
  \resizebox{1.0\textwidth}{!}{
  \begin{tabular}{r|lc|lr|lr}
    \toprule
    \multicolumn{1}{r}{\textbf{ID}} &
    \multicolumn{1}{l}{\textbf{Assistant}} &
    \multicolumn{1}{c|}{\textbf{Release}} &
    \multicolumn{1}{l}{\textbf{Wake Word(s)}} &
    \multicolumn{1}{c|}{\textbf{Lang.\textsuperscript{\scriptsize{$\dagger$}}}} &
    \multicolumn{1}{l}{\textbf{Smart Speaker}} &
    \multicolumn{1}{c}{\textbf{SW. Version}}\\
    \midrule
    VA1  & Amazon: Alexa        & 2014 & \emph{Alexa}                & en\_us, de\_de & Amazon: Echo Dot (v3)           & 392 657 0628   \\ % Sep. 2018
    VA2  & Amazon: Alexa        & 2014 & \emph{Computer}             & en\_us, de\_de & Amazon: Echo Dot (v3)           & 392 657 0628   \\ % Sep. 2018
    VA3  & Amazon: Alexa        & 2014 & \emph{Echo}                 & en\_us, de\_de & Amazon: Echo Dot (v3)           & 392 657 0628   \\ % Sep. 2018
    VA4  & Amazon: Alexa        & 2014 & \emph{Amazon}               & en\_us, de\_de & Amazon: Echo Dot (v3)           & 392 657 0628   \\ % Sep. 2018
    VA5  & Google: Assistant    & 2012 & \emph{OK/Hey Google}        & en\_us, de\_de & Google: Home Mini               & 191 160        \\ % Nov. 2017
    VA6  & Apple: Siri          & 2011 & \emph{Hey Siri}             & en\_us, de\_de & Apple: HomePod                  & 13.4.8         \\ % Feb. 2018
    VA7  & Microsoft: Cortana   & 2014 & \emph{Hey/- Cortana}        & en\_us     & Harman Kardon: Invoke               & 11.1842.0      \\ % Oct. 2017
    VA8  & Xiaomi: Xiao AI      & 2017 & \emph{Xiǎo ài tóngxué}      & zh\_cn     & Xiaomi: Mi AI Speaker               & 1.52.4         \\ % Jul. 2017
    VA9  & Tencent: Xiaowei     & 2017 & \emph{Jiǔsì'èr líng}        & zh\_cn     & Tencent: Tīngtīng TS-T1             & 3.5.0.025      \\ % Apr. 2018
    VA10 & Baidu: DuerOS        & 2015 & \emph{Xiǎo dù xiǎo dù}      & zh\_cn     & Baidu: NV6101 (1C)                  & 1.34.5         \\ % Mar. 2018
    VA11 & SoundHound: Houndify & 2015 & \emph{Hallo/Hey/Hi Magenta} & de\_de     & Deutsche Telekom: Magenta Speaker   & 1.1.2          \\ % Sep. 2019
    \bottomrule
  \end{tabular}}\label{tab:va}
  \begin{flushleft}
  \small
  $\dagger$: In our experiments, we only considered English (US), German (DE), and Standard Chinese (ZH).\\
  \end{flushleft}
  \vspace{-1.5em}
\end{table*}

\subsection{Evaluated Smart Speakers}\label{sec:selected_assistants}
In our experiments, we evaluate 11~smart speakers as listed in Table~\ref{tab:va}. The smart speakers have been selected based on their market shares and availability~\cite{canalys-19-smart-speaker-market-share, lyons-20-alexa-marketshare}. In the following, with the term \emph{smart speaker}, we refer to the hardware component. At the same time, we use the term \emph{voice assistant} to refer to cloud-assisted ASR and the conversational intelligence built into the speaker.

Since its introduction in 2014, the Amazon Echo is one of the most popular speakers. It enables users to choose between four different wake words (``Alexa,'' ``Computer,'' ``Echo,'' and ``Amazon''). In our experiments, we used four Echo Dot (3rd~Gen.) and configured each to a different wake word. Similarly, for the Google Assistant, we used a Home Mini speaker, which listens to the wake words ``OK~Google'' and ``Hey~Google.'' From Apple, we evaluated a HomePod speaker with ``Hey~Siri'' as its wake word. To test Microsoft's Cortana, we bought the official Invoke smart speaker developed by Harman Kardon that recognizes ``Cortana'' and ``Hey~Cortana.'' 

Moreover, we expanded the set by including non-English~(US) speaking assistants from Europe and Asia. We bought three Standard Chinese~(ZH) and one German~(DE) speaking smart speaker. The Xiaomi speaker listens to ``Xiǎo~ài~tóngxué'' (\begin{CJK*}{UTF8}{gbsn}小爱同学\end{CJK*}), which literately translates to ``little classmate.'' The Tencent speaker listens to ``Jiǔsì'èr~líng'' (\begin{CJK*}{UTF8}{gbsn}九四二零\end{CJK*}), which literately translates to the digit sequence 9-4-2-0. The wake word is a phonetic replacement of ``Jiùshì ài nǐ,'' which translates to ``just love you.'' The Baidu speaker listens to ``Xiǎo~dù xiǎo~dù'' (\begin{CJK*}{UTF8}{gbsn}小度小度\end{CJK*}), which literately translates to ``small degree,'' but is related to the smart device product line Xiaodu (little ``du'' as in Bai\emph{du}). Finally, we ordered the Magenta Speaker from the German telecommunications operator Deutsche Telekom, which listens to ``Hallo,'' ``Hey,'' and ``Hi~Magenta.'' In this case, magenta refers to a product line and also represents the company's primary brand color. Deutsche Telekom has not developed the voice assistant on their own. Instead, they chose to integrate a third-party white-label solution developed by SoundHound~\cite{kinsella-19-magenta-houndify}. While the speaker also allows accessing Amazon Alexa, we have not enabled this feature for our measurements. The Magenta Speaker is technically identical to the Djingo speaker~\cite{orange-19-djingo}, which was co-developed by the French operator Orange.

\subsection{Evaluated Datasets}\label{sec:datasets} In the following, we provide an overview of the datasets used to evaluate the prevalence of accidental triggers. We included media to resemble content, which is most likely played in a typical US household to simulate an environment with ambient noise from external sources such as TVs~\cite{raju-18-robust-wake-word}. Moreover, we considered professional audio datasets used by the machine learning community.

\paragraph{TV Shows} The first category of media is TV shows. We considered a variety of different genres to be most representative. Our list comprises popular shows from the last 10~years and includes animated and a family sitcom, a fantasy drama, and a political thriller. Our English (US) TV show dataset includes \emph{Game of Thrones}, \emph{House of Cards}, \emph{Modern Family}, \emph{New Girl}, and \emph{The Simpsons}.

\paragraph{News} The second category is newscasts. As newscasts tend to be repetitive, we used one broadcast per day and television network only. The analyzed time frame covers news broadcast between August and October 2019. Our English (US) newscasts dataset includes \emph{ABC World News}, \emph{CBS Evening News}, \emph{NBC Nightly News}, and \emph{PBS NewsHour}.

\paragraph{Professional Datasets} The third category is professional audio datasets. Due to the costly process of collecting appropriate training datasets and the accessibility of plenty and well-analyzed datasets, we considered professional audio datasets commonly used by the speech recognition community.
To verify whether voice assistants are trained on such datasets, we included the following English (US) datasets:
\begin{compactitem}
\item \emph{LibriSpeech}~\cite{panayotov-15-librispeech}:
An audio dataset created by volunteers who read and record public domain texts to create audiobooks. It contains $1,000$ hours of speech. The corpus has been built in 2015 and is publicly available; it is a widely used benchmark for automatic speech recognition.

\item \emph{Mozilla Common Voice}~\cite{mozilla-19-commonvoice}: The dataset is based on an ongoing crowdsourcing project headed by Mozilla to create a free speech database. At the time of writing, the project includes a collection of $48$ languages. Our English (US) version of the dataset contains $1,200$~hours of speech and has been downloaded in August 2019. As neither the environment nor the equipment for the audio recordings is controlled, the quality of the recordings differs widely.
\item \emph{Wall Street Journal}~\cite{paul-92-wsj}: A corpus developed to support research on large-vocabulary, continuous speech recognition systems containing read English text. The dataset was recorded in 1993 in a controlled environment and comprises $400$~hours of speech.
\item \emph{CHiME}: The CHiME (Computational Hearing in Multisource Environments) dataset is intended to train models to recognize speech from recordings made by distant microphones in noisy environments. The \emph{5th} CHiME challenge dataset includes recordings from a group dinner of four participants each, with two acting as hosts and two as guests~\cite{barker-18-chime}. Audio signals were recorded at 20 parties, each in a different home, via six Kinect microphone arrays and four binaural microphone pairs. This dataset thus provides multi-channel recordings of highly realistic, distant-talking speech with natural background noise. In total, the dataset consists of 50 hours of recording time.
\end{compactitem}

\paragraph{Noise} We used noise recordings as a special category to test the sensitivity of the voice assistants against audio data other than speech. For this purpose, we used the noise partition of the \emph{MUSAN}~dataset~\cite{snyder-15-musan}, containing approximately $6$~hours of many kinds of environmental noise (excluding speech and music data).

\paragraph{Non-English Media} To test for linguistic differences, e.\,g., biases between different languages, we tested one Standard Chinese (ZH) and four German (DE) TV shows. We analyzed the Chinese TV show \emph{All Is Well} and German-dubbed version of the TV show \emph{Modern Family} for easy comparison. Additionally, we tested the German-dubbed version of \emph{The Big Bang Theory}, as well as, \emph{Polizeiruf 110} and \emph{Tatort} as examples for undubbed German TV shows. Moreover, we evaluated three shorter (12 hours) samples of the Chinese newscast \emph{CCTV Xinwen Lianbo} and the German newscasts \emph{ARD Tagesschau} and \emph{ZDF Heute Journal}.

\paragraph{Female vs.\ Male Speakers} To explore potential gender biases in accidental triggers of voice assistants, we included two sets of randomly chosen voice data from the aforementioned \emph{LibriSpeech} dataset. Every set consisted of a female and a male $24$ hour sample. Every sample was built from multiple $20$ minutes sequences, which themselves were made of 100 different $12$ seconds audio snippets.

%% file: prevalence.tex
\section{Prevalence of Accidental Triggers}\label{sec:prevalence}

Based on the datasets described above, we now explore the prevalence of accidental triggers in various media such as TV shows, newscasts, and professional audio datasets.

\subsection{Approach}
We start by describing our technical setup to measure the prevalence of accidental triggers across 11~smart speakers (cf. Section~\ref{sec:selected_assistants}) using $24$ hour samples of various datasets (cf. Section~\ref{sec:datasets}). The basic idea is to simulate a common living room-like scenario, where a smart speaker is in close proximity to an external audio source like a TV~\cite{raju-18-robust-wake-word,sigtia-20-apple-trigger-detection-tv,sigtia-20-apple-speaker-detection-tv}.

\begin{figure}[tbp]
  \centering
  \includegraphics[width=0.85\columnwidth]{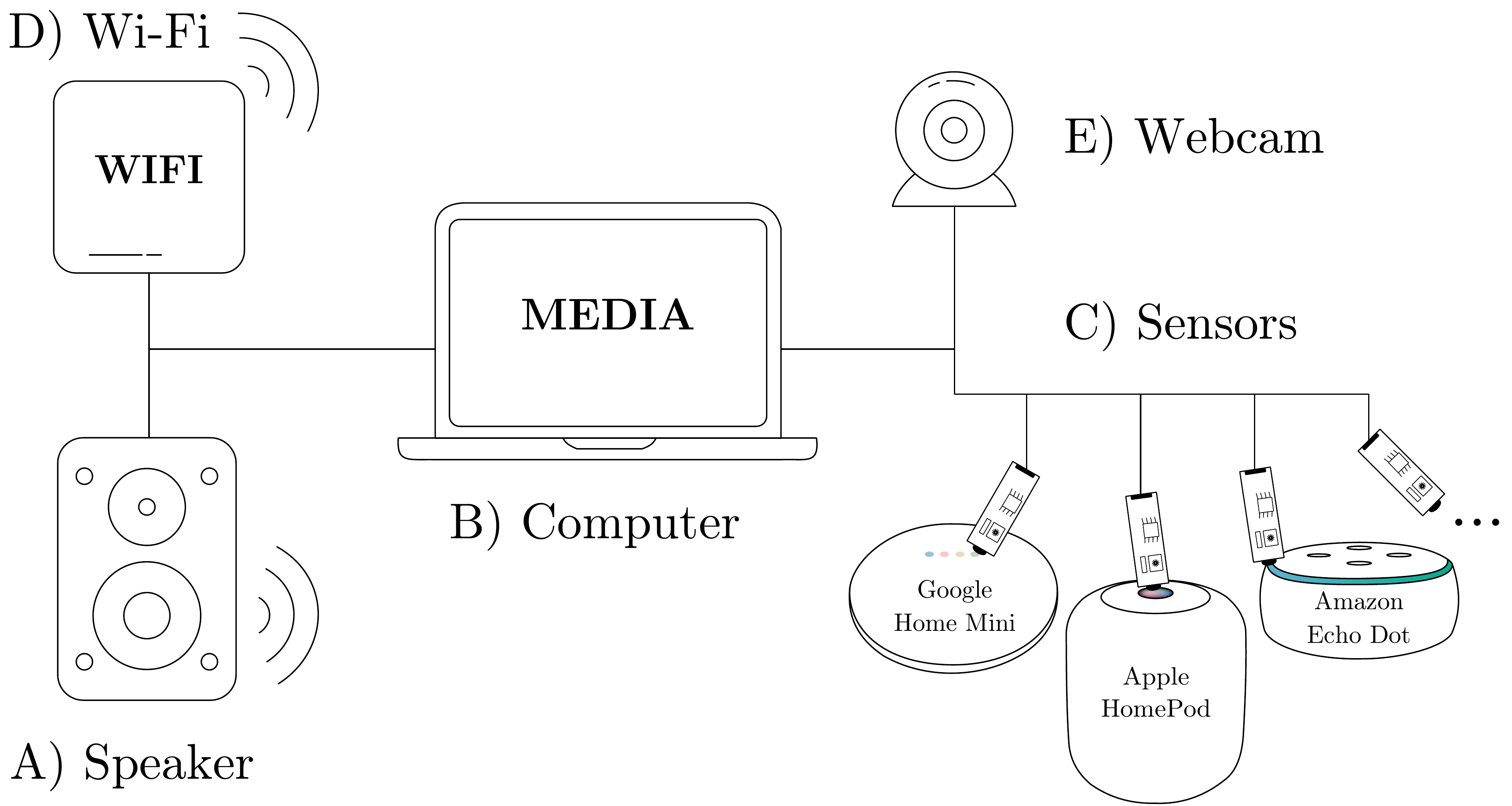}
  \caption{Setup: A loudspeaker~(A) is playing media files from a computer~(B). The LED activity indicators of a group of smart speakers are monitored using light sensors~(C). All speakers are connected to the Internet over Wi-Fi~(D). A webcam (E) is used to record a video of each measurement.}\label{fig:measrument_setup}
\end{figure}

\subsubsection{Measurement Setup}

\begin{figure}[tb]
  \centering
  \includegraphics[width=0.9\columnwidth]{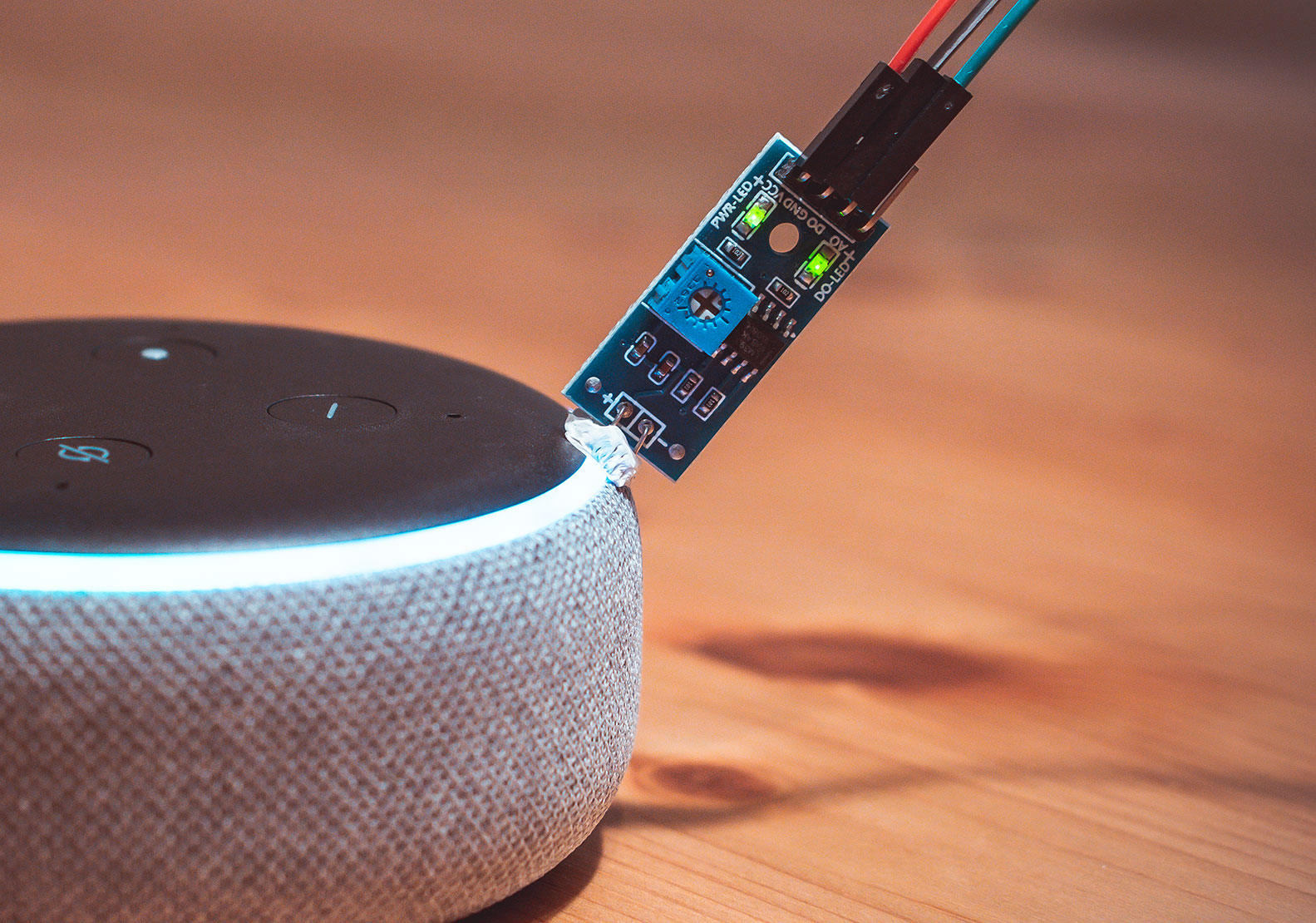}
  \caption{Photoresistor attached to the LED indicator of a smart speaker. The sensitivity of the sensor can be adjusted via a potentiometer. Any activity is recognized and logged.}\label{fig:photoresistor}
\end{figure}

\paragraph{Hardware}
The measurement setup consists of five components, as depicted in Figure~\ref{fig:measrument_setup}. To rule out any external interference, all experiments are conducted in a sound-proof chamber. We positioned 11~smart speakers at a distance of approx. $1$~meter to a loudspeaker~(A) and play media files from a computer~(B). To detect any activity of the smart speakers, we attach photoresistors~(C) (i.\,e., light sensors) on the LED activity indicator of each speaker, as one can see in Figure~\ref{fig:photoresistor}. In the case of any voice assistant activity, the light sensor detects the spontaneous change in brightness and emits a signal to the computer~(B). To prevent interference from external light sources, the photoresistors are covered by a small snippet of reusable adhesive tape.
All smart speakers are connected to the Internet using a WiFi network~(D). During all measurements, we record network traces using \texttt{tcpdump} to be able to analyze their activity on a network level. To verify the measurement results, we record a video of each measurement via a webcam with a built-in microphone~(E). The entire setup is connected to a network-controllable power socket that we use to power cycle the speakers in case of failures or non-responsiveness.

\begin{figure*}[!htbp]
  \centering
  \includegraphics[width=0.9\textwidth]{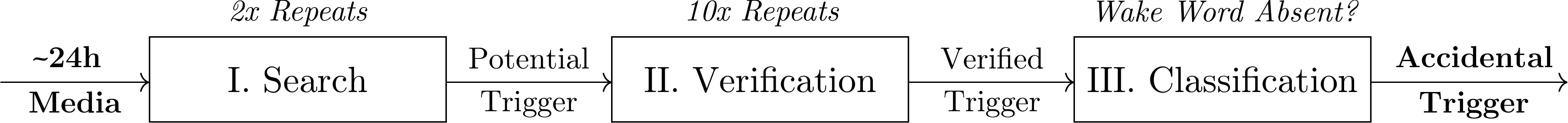}
  \caption{Trigger Detection Workflow: Every approx. 24-hour dataset is played twice. Subsequently, the existence of every \emph{potential} trigger is confirmed. Finally, every \emph{verified} trigger is classified as \emph{accidental}, if the wake word or a related word is not present in the identified scene.}\label{fig:workflow}
\end{figure*}

\paragraph{Software}
To verify the functionality and responsiveness of the measurement setup, we periodically play a test signal comprising the wake word of each voice assistant. The test signal consists of the wake word~(\eg ``Alexa'') and the stop word~(\eg ``Stop'') of each voice assistant (in its configured language) and a small pause in-between. Overall, the test signal for all 11~speakers is approximately 2m~30s long. During the measurements, we verify that each voice assistant triggers to its respective test signal. In the case of no response, multiple or prolonged responses, all voice assistants are automatically rebooted and rechecked. As a side effect, the test signal ensures that each assistant stops any previous activity (like playing music or telling a joke). That might have been accidentally triggered by a previous measurement run. Using this setup, we obtain a highly reliable and fully automated accidental trigger search system.

\subsubsection{Trigger Detection}\label{sec:trigger-detection}
The process of measuring the prevalence of accidental triggers consists of three parts, as depicted in Figure~\ref{fig:workflow}. First, a 24-hour search is executed twice per dataset. Second, a ten-fold verification of a \emph{potential} trigger is done to confirm the existence of the trigger and measure its reproducibility. Third, a manual classification of \emph{verified} triggers is performed to ensure the absence of the wake word or related words. In the following, we describe these steps in more detail.

\noindent \textbf{I. Search:} In a first step, we prepare a 24-hour audio sample consisting of multiple episodes/broadcasts of approximately $20$~minutes each (with slightly different lengths depending on the source material) for each of the datasets introduced in Section~\ref{sec:datasets}. We play each of the 24-hour samples twice and log any smart speaker LED activity as an indicator for a potential trigger. The logfile includes a timestamp, the currently played media, the playback progress in seconds, and the triggered smart speaker's name.

We played each audio file twice due to some changes in results that were observed when we played the same sample multiple times. These changes do not come as a great surprise, given that they are due to the internal framing of the recorded audio. Therefore, each time one plays the same audio file, the system will get a slightly different signal, with slightly shifted windows and possibly small changes in the additive noise that cannot be fully prevented but is (strongly) damped in our test environment. Also, there may be further indeterminacies up in the chain of trigger processing, as was also noted by others~\cite{kumar-18-skillsquatting,dubois-20-characterizing-misactivations}.

\noindent \textbf{II. Verification:} In a second step, we extract a list of \emph{potential} triggers from the logfile and verify these triggers by replaying a $10$-second snippet containing the identified scene. From the potential trigger location within the media, i.\,e., the playback progress when the trigger occurred, we rewind 7~seconds and replay the scene until 3~seconds after the documented trigger location. This playback is repeated ten times to confirm the existence and to measure the reproducibility of the trigger.

\noindent \textbf{III. Classification}:
In a third step, every verified trigger is classified by reviewing a 30-second snippet of the webcam recording at the time of the trigger.Here, two independently working reviewers need to confirm the accidental trigger by verifying the correct wake word's absence. If a trigger is caused by the respective wake word or a related word such as Alexander (``Alexa''), computerized (``Computer''), echoing (``Echo''), Amazonian (``Amazon''), etc., we discard the trigger and exclude it from further analysis. Where available, the analysis is assisted by the transcriptions/subtitles of the respective dataset.

To determine the approximate distribution between \emph{local} and \emph{cloud}-based triggers, we expand our classification step. Instead of only determining the mere presence of the wake word or a related word, two members of our team also classify the triggers into local or local + cloud triggers. As noted in Section~\ref{sec:privacy-settings}, not all smart speaker vendors provide access or report local triggers in their voice assistant interaction history. Thus, we use the internal processes, especially the LED timings and patterns, to classify triggers.  The heuristic for that classification is based on the time the LED indicator of the speaker remains on. Based on preliminary tests, we choose a conservative threshold of $2$~seconds of speaker activity to classify the trigger as local + cloud. Moreover, we use voice responses and certain LED patterns as obvious signals for a local + cloud trigger. The inter-rater reliability between our reviewers, measured by Cohen's kappa, is $\kappa \geq 0.89$ across all evaluated datasets.

\subsection{Results}\label{sec:prevalence-results}
An overview of our results can be found in Table~\ref{tab:prevalence}. We report the absolute counts of observed accidental triggers and actual instances of spoken wake words.

\paragraph{Comparison Across Speakers}
Looking at the four \mbox{VA1-4} Amazon Echo wake words, we can see that ``Amazon'' ($57$) and ``Echo'' ($43$) trigger less often than ``Alexa'' ($99$) and ``Computer'' ($77$).
Moreover, we observe that the VA5~Google Home ($16$) and the VA6~Apple HomePod ($8$) seem to be the most robust speakers of all English (US) speakers across all played datasets, and we discuss potential reasons for that in Section~\ref{sec:wake-word-discussion}.
Another noteworthy observation is that VA7~Microsoft Cortana triggered far more often ($197$) than the other speakers across all kinds of audio data.

From a qualitative perspective, the identified triggers are often confusions with similar-sounding words or sequences, such as, ``a lesson'' (Alexa), ``commuter'' (Computer), ``OK, cool'' (OK Google), ``a city'' (Hey Siri). Another category are proper names that are unknown or likely infrequently included in the training data. Examples include names of persons and states such as ``Peter'' and ``Utah'' (Computer), ``Eddard'' (Echo), ``Montana'' (Cortana), but also uncommon old English phrases such as ``Alas!'' (Alexa). Finally, we observed a few cases of triggers that include fictional language (\emph{Dothraki}) or unintelligible language (\emph{gibberish}) and two occasions of \emph{non-speech} accidental triggers: A ringing phone triggering ``Amazon'' in the TV show \emph{New Girl} and a honk made by a car horn triggering ``Alexa'' in the TV show \emph{The Simpsons}.

%%%%%%%%%%%%%%%%%%%%%%%
%%% MAIN RESULTS TABLE
%%%%%%%%%%%%%%%%%%%%%%%
\begin{table*}[htbp]
  \centering
  \caption{Prevalence of Accidental Triggers and Wake Words.}
  \smallskip
  \resizebox{1.0\textwidth}{!}{
    
    \begin{tabular}{rc|cc|cc|cc|cc|cc|cc|cc||cc|cc|cc|cc}
    \toprule
    \multicolumn{1}{r}{} & & \multicolumn{2}{c|}{\multirow{2}[2]{*}{\textbf{Alexa}}} & \multicolumn{2}{c|}{\multirow{2}[2]{*}{\textbf{Computer}}} & \multicolumn{2}{c|}{\multirow{2}[2]{*}{\textbf{Echo}}} & \multicolumn{2}{c|}{\multirow{2}[2]{*}{\textbf{Amazon}}} & \multicolumn{2}{c|}{\textbf{Ok}} & \multicolumn{2}{c|}{\textbf{Hey}} & \multicolumn{2}{c||}{\textbf{Hey}} & \multicolumn{2}{c|}{\textbf{Xiǎo ài}} & \multicolumn{2}{c|}{\textbf{Jiǔsì'}} & \multicolumn{2}{c|}{\textbf{Xiǎo dù}} & \multicolumn{2}{c}{\textbf{Hallo}} \\
    \multicolumn{1}{r}{} & & \multicolumn{2}{c|}{} & \multicolumn{2}{c|}{} & \multicolumn{2}{c|}{} & \multicolumn{2}{c|}{} & \multicolumn{2}{c|}{\textbf{Google}} & \multicolumn{2}{c|}{\textbf{Siri}} & \multicolumn{2}{c||}{\textbf{Cortana}} & \multicolumn{2}{c|}{\textbf{tóngxué}} & \multicolumn{2}{c|}{\textbf{èr líng}} & \multicolumn{2}{c|}{\textbf{xiǎo dù}} & \multicolumn{2}{c}{\textbf{Magenta}} \\
\cmidrule{3-24}    \multicolumn{1}{r}{} & & A     & W     & A     & W     & A     & W     & A     & W     & A     & W     & A     & W     & A     & W     & A     & W     & A     & W     & A     & W     & A     & W \\
\cmidrule{3-24}    \multicolumn{1}{r}{} & & \multicolumn{2}{c|}{\textbf{en\_us}} & \multicolumn{2}{c|}{\textbf{en\_us}} & \multicolumn{2}{c|}{\textbf{en\_us}} & \multicolumn{2}{c|}{\textbf{en\_us}} & \multicolumn{2}{c|}{\textbf{en\_us}} & \multicolumn{2}{c|}{\textbf{en\_us}} & \multicolumn{2}{c||}{\textbf{en\_us}} & \multicolumn{2}{c|}{\cellcolor{gray!10}zh\_cn} & \multicolumn{2}{c|}{\cellcolor{gray!10}zh\_cn} & \multicolumn{2}{c|}{\cellcolor{gray!10}zh\_cn} & \multicolumn{2}{c}{\cellcolor{gray!10}de\_de} \\

    \midrule
    \textbf{TV Shows} & \textbf{Time} & \textbf{31} & \textbf{0} & \textbf{31} & \textbf{6} & \textbf{18} & \textbf{2} & \textbf{38} & \textbf{2} & \textbf{3} & \textbf{0} & \textbf{2} & \textbf{0} & \textbf{94} & \textbf{0} & \cellcolor{gray!10}\textbf{1} & \cellcolor{gray!10}\textbf{0} & \cellcolor{gray!10}\textbf{7} & \cellcolor{gray!10}\textbf{0} & \cellcolor{gray!10}\textbf{0} & \cellcolor{gray!10}\textbf{0} & \cellcolor{gray!10}\textbf{3} & \cellcolor{gray!10}\textbf{0} \\
    \midrule
    Game of Thrones & 24h & 6     & -     & 6     & -     & 5     & -     & 3     & -     & -     & -     & -     & -     & 14    & -     & \cellcolor{gray!10}1     & \cellcolor{gray!10}-     & \cellcolor{gray!10}1     & \cellcolor{gray!10}-     & \cellcolor{gray!10}-     & \cellcolor{gray!10}-     & \cellcolor{gray!10}1     & \cellcolor{gray!10}- \\
    House of Cards & 24h & 2     & -     & 11    & -     & 2     & 2     & 15    & -     & -     & -     & 1     & -     & 14    & -     & \cellcolor{gray!10}-     & \cellcolor{gray!10}-     & \cellcolor{gray!10}3     & \cellcolor{gray!10}-     & \cellcolor{gray!10}-     & \cellcolor{gray!10}-     & \cellcolor{gray!10}1     & \cellcolor{gray!10}- \\
    Modern Family & 24h & 6     & -     & 9     & 4     & 4     & -     & 12    & 1     & 1     & -     & 1     & -     & 23    & -     & \cellcolor{gray!10}-     & \cellcolor{gray!10}-     & \cellcolor{gray!10}1     & \cellcolor{gray!10}-     & \cellcolor{gray!10}-     & \cellcolor{gray!10}-     & \cellcolor{gray!10}1     & \cellcolor{gray!10}- \\
    New Girl & 24h & 4     & -     & 5     & 1     & 4     & -     & 6     & -     & 2     & -     & -     & -     & 29    & -     & \cellcolor{gray!10}-     & \cellcolor{gray!10}-     & \cellcolor{gray!10}1     & \cellcolor{gray!10}-     & \cellcolor{gray!10}-     & \cellcolor{gray!10}-     & \cellcolor{gray!10}-     & \cellcolor{gray!10}- \\
    The Simpsons & 24h & 13    & -     & -     & 1     & 3     & -     & 2     & 1     & -     & -     & -     & -     & 14    & -     & \cellcolor{gray!10}-     & \cellcolor{gray!10}-     & \cellcolor{gray!10}1     & \cellcolor{gray!10}-     & \cellcolor{gray!10}-     & \cellcolor{gray!10}-     & \cellcolor{gray!10}-     & \cellcolor{gray!10}- \\
    \midrule
    \textbf{News} & \textbf{Time} & \textbf{22} & \textbf{5} & \textbf{9} & \textbf{2} & \textbf{4} & \textbf{4} & \textbf{12} & \textbf{62} & \textbf{2} & \textbf{0} & \textbf{4} & \textbf{2} & \textbf{44} & \textbf{0} & \cellcolor{gray!10}\textbf{1} & \cellcolor{gray!10}\textbf{0} & \cellcolor{gray!10}\textbf{4} & \cellcolor{gray!10}\textbf{0} & \cellcolor{gray!10}\textbf{0} & \cellcolor{gray!10}\textbf{0} & \cellcolor{gray!10}\textbf{1} & \cellcolor{gray!10}\textbf{0} \\
    \midrule
    ABC World News  & 24h & -     & -     & 3     & -     & -     & -     & 2     & 9     & 1     & -     & 1     & -     & 11    & -     & \cellcolor{gray!10}-     & \cellcolor{gray!10}-     & \cellcolor{gray!10}1     & \cellcolor{gray!10}-     & \cellcolor{gray!10}-     & \cellcolor{gray!10}-     & \cellcolor{gray!10}-     & \cellcolor{gray!10}- \\
    CBS Evening News & 24h & 12    & 1     & 1     & 1     & -     & -     & 7     & 24    & -     & -     & -     & -     & 13    & -     & \cellcolor{gray!10}-     & \cellcolor{gray!10}-     & \cellcolor{gray!10}1     & \cellcolor{gray!10}-     & \cellcolor{gray!10}-     & \cellcolor{gray!10}-     & \cellcolor{gray!10}1     & \cellcolor{gray!10}- \\
    NBC Nightly News & 24h & 2     & 4     & -     & -     & 2     & 1     & -     & 23    & 1     & -     & 2     & 2     & 6     & -     & \cellcolor{gray!10}-     & \cellcolor{gray!10}-     & \cellcolor{gray!10}2     & \cellcolor{gray!10}-     & \cellcolor{gray!10}-     & \cellcolor{gray!10}-     & \cellcolor{gray!10}-     & \cellcolor{gray!10}- \\
    PBS NewsHour & 24h & 8     & -     & 5     & 1     & 2     & 3     & 3     & 6     & -     & -     & 1     & -     & 14    & -     & \cellcolor{gray!10}1     & \cellcolor{gray!10}-     & \cellcolor{gray!10}-     & \cellcolor{gray!10}-     & \cellcolor{gray!10}-     & \cellcolor{gray!10}-     & \cellcolor{gray!10}-     & \cellcolor{gray!10}- \\
    \midrule
    \textbf{Professional} & \textbf{Time} & \textbf{46} & \textbf{1} & \textbf{37} & \textbf{32} & \textbf{21} & \textbf{3} & \textbf{7} & \textbf{1} & \textbf{11} & \textbf{0} & \textbf{2} & \textbf{0} & \textbf{59} & \textbf{0} & \cellcolor{gray!10}\textbf{2} & \cellcolor{gray!10}\textbf{0} & \cellcolor{gray!10}\textbf{3} & \cellcolor{gray!10}\textbf{0} & \cellcolor{gray!10}\textbf{0} & \cellcolor{gray!10}\textbf{0} & \cellcolor{gray!10}\textbf{1} & \cellcolor{gray!10}\textbf{0} \\
    \midrule
    LibriSpeech & 24h & 14    & -     & 9     & -     & 6     & 2     & 5     & -     & -     & -     & -     & -     & 17    & -     & \cellcolor{gray!10}-     & \cellcolor{gray!10}-     & \cellcolor{gray!10}-     & \cellcolor{gray!10}-     & \cellcolor{gray!10}-     & \cellcolor{gray!10}-     & \cellcolor{gray!10}-     & \cellcolor{gray!10}- \\
    Moz. CommonVoice & 24h & 10    & 1     & 21    & 5     & 14    & 1     & 2     & 1     & 11    & -     & 2     & -     & 18    & -     & \cellcolor{gray!10}1     & \cellcolor{gray!10}-     & \cellcolor{gray!10}1     & \cellcolor{gray!10}-     & \cellcolor{gray!10}-     & \cellcolor{gray!10}-     & \cellcolor{gray!10}-     & \cellcolor{gray!10}- \\
    WallStreetJournal & 24h & 22    & -     & 7     & 27    & 1     & -     & -     & -     & -     & -     & -     & -     & 24    & -     & \cellcolor{gray!10}1     & \cellcolor{gray!10}-     & \cellcolor{gray!10}2     & \cellcolor{gray!10}-     & \cellcolor{gray!10}-     & \cellcolor{gray!10}-     & \cellcolor{gray!10}1     & \cellcolor{gray!10}- \\
    CHiME & 24h & 1    & -     & 3     & 3    & -     & -     & 10     & 2     & 7     & -     & 1     & -     & 1    & -     & \cellcolor{gray!10}-     & \cellcolor{gray!10}-     & \cellcolor{gray!10}1     & \cellcolor{gray!10}-     & \cellcolor{gray!10}-     & \cellcolor{gray!10}-     & \cellcolor{gray!10}-     & \cellcolor{gray!10}- \\
    \midrule
    \textbf{Sum} & 13d & \textbf{100} & \textbf{6} & \textbf{80} & \textbf{43} & \textbf{43} & \textbf{9} & \textbf{67} & \textbf{67} & \textbf{23} & \textbf{0} & \textbf{9} & \textbf{2} & \textbf{198} & \textbf{0} & \cellcolor{gray!10}\textbf{4} & \cellcolor{gray!10}\textbf{0} & \cellcolor{gray!10}\textbf{15} & \cellcolor{gray!10}\textbf{0} & \cellcolor{gray!10}\textbf{0} & \cellcolor{gray!10}\textbf{0} & \cellcolor{gray!10}\textbf{5} & \cellcolor{gray!10}\textbf{0} \\
    \bottomrule
    \end{tabular}}
    \label{tab:prevalence}
    \begin{flushleft}
    \small \vspace{-0.5em}
    $A$: Accidental triggers; $W$: Wake word said; Gray cells: Mismatch between played audio and wake word model language.\\
    \end{flushleft}
    \vspace{-1.5em}
\end{table*}%

\paragraph{Comparison Across Datasets}
When comparing across datasets, one must keep in mind that the total playback time differs across categories. While every dataset (i.\,e., every row in the table) consisted of 24 hours of audio data, the number of datasets per category differs.

In general, we cannot observe any noteworthy differences in accidental triggers (A) across the three dataset categories. In contrast, if we have a look at the cases where the wake word was actually said (W), we see that this was very often the case for ``Computer'' in the professional Wall Street Journal dataset caused by an article about the computer hardware company IBM and for ``Amazon'' across the news datasets. In this case, the $62$~instances of ``Amazon'' referred $13$~times to the 2019 Amazon rainforest wildfires and $49$~times to the company.

If we look at the professional datasets, the number of triggers is within the same range or even increases compared to TV shows and news. As such, we have not found a speaker that triggered less often, because it might have been specifically trained on one of the professional datasets. In contrast to the other professional audio datasets, the CHiME dataset consists of recordings of group dinner scenarios resulting in comparatively less spoken words, explaining the overall lower number of accidental activations.
Not presented in Table~\ref{tab:prevalence} is the \emph{MUSAN} noise dataset, because we have not observed any triggers across the different speakers. This suggests that accidental triggers are less likely to occur for non-speech audio signals.

\begin{figure}[tbp]
  \centering
  \includegraphics[width=0.9\columnwidth]{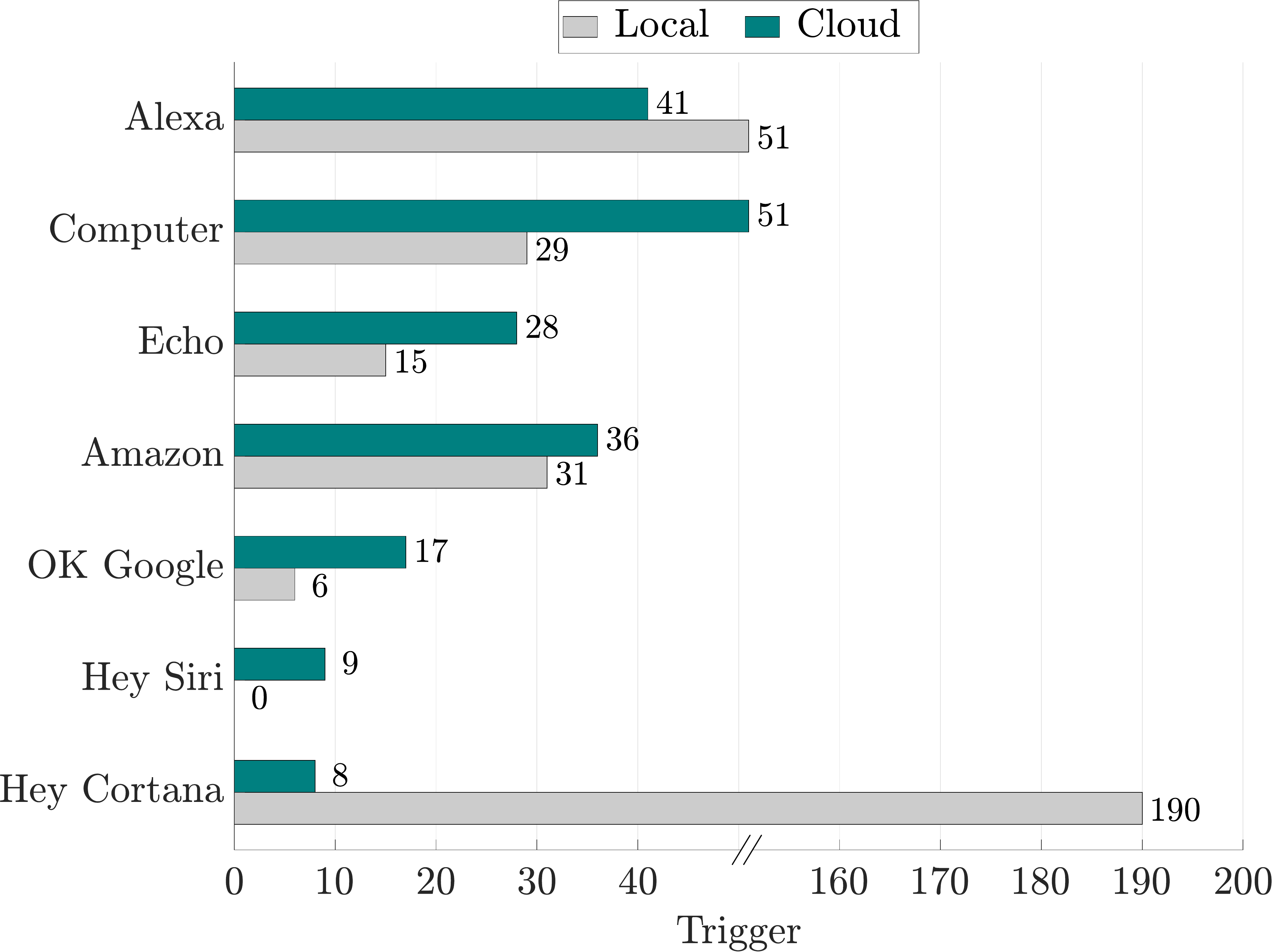}
  \caption{Number of accidental triggers that are incorrectly recognized by the local and the cloud-based ASR engine.}
  \label{fig:local-vs-cloud}
  \vspace{-1em}
\end{figure}

\paragraph{Comparison Between Local and Cloud-Based Triggers}\label{sec:local-vs-cloud}
An overview of the distribution can be seen in Figure~\ref{fig:local-vs-cloud}. Depending on the wake word, we find that the cloud ASR engine also misrecognizes about half of our accidental triggers. Fortunately for Cortana, only a small number of triggers (8 out of 197) are able to trick Microsoft's cloud verification.

\paragraph{Comparison Between Female and Male Speakers}\label{sec:female-vs-male}
We performed an experiment designed to study a potential model bias in smart speakers, a common problem for machine learning systems~\cite{dixon-18-measuring,tatman-17-gender,kiritchenko-18-examining}. Fortunately, across our tested datasets, we cannot find any noteworthy difference in the number of accidental triggers for female and male speakers; no bias can be observed in our experiments. The detailed numbers are shown in Table~\ref{tab:gender}. 

%%%%%%%%%%%%%%%%%%%%%%%
%%% GENDER TABLE
%%%%%%%%%%%%%%%%%%%%%%%
\begin{table*}[tbp]
  \centering
  \caption{Differences Between Female and Male Speakers.}
  \smallskip
  \resizebox{1.0\textwidth}{!}{
    
    \begin{tabular}{rc|cc|cc|cc|cc|cc|cc|cc||cc|cc|cc|cc}
    \toprule
    \multicolumn{1}{r}{} & & \multicolumn{2}{c|}{\multirow{2}[2]{*}{\textbf{Alexa}}} & \multicolumn{2}{c|}{\multirow{2}[2]{*}{\textbf{Computer}}} & \multicolumn{2}{c|}{\multirow{2}[2]{*}{\textbf{Echo}}} & \multicolumn{2}{c|}{\multirow{2}[2]{*}{\textbf{Amazon}}} & \multicolumn{2}{c|}{\textbf{Ok}} & \multicolumn{2}{c|}{\textbf{Hey}} & \multicolumn{2}{c||}{\textbf{Hey}} & \multicolumn{2}{c|}{\textbf{Xiǎo ài}} & \multicolumn{2}{c|}{\textbf{Jiǔsì'}} & \multicolumn{2}{c|}{\textbf{Xiǎo dù}} & \multicolumn{2}{c}{\textbf{Hallo}} \\
    \multicolumn{1}{r}{} & & \multicolumn{2}{c|}{} & \multicolumn{2}{c|}{} & \multicolumn{2}{c|}{} & \multicolumn{2}{c|}{} & \multicolumn{2}{c|}{\textbf{Google}} & \multicolumn{2}{c|}{\textbf{Siri}} & \multicolumn{2}{c||}{\textbf{Cortana}} & \multicolumn{2}{c|}{\textbf{tóngxué}} & \multicolumn{2}{c|}{\textbf{èr líng}} & \multicolumn{2}{c|}{\textbf{xiǎo dù}} & \multicolumn{2}{c}{\textbf{Magenta}} \\
    \cmidrule{3-24}    \multicolumn{1}{r}{} & & A     & W     & A     & W     & A     & W     & A     & W     & A     & W     & A     & W     & A     & W     & A     & W     & A     & W     & A     & W     & A     & W \\
    
\cmidrule{3-24}    \multicolumn{1}{r}{} & & \multicolumn{2}{c|}{\textbf{en\_us}} & \multicolumn{2}{c|}{\textbf{en\_us}} & \multicolumn{2}{c|}{\textbf{en\_us}} & \multicolumn{2}{c|}{\textbf{en\_us}} & \multicolumn{2}{c|}{\textbf{en\_us}} & \multicolumn{2}{c|}{\textbf{en\_us}} & \multicolumn{2}{c||}{\textbf{en\_us}} & \multicolumn{2}{c|}{\cellcolor{gray!10}zh\_cn} & \multicolumn{2}{c|}{\cellcolor{gray!10}zh\_cn} & \multicolumn{2}{c|}{\cellcolor{gray!10}zh\_cn} & \multicolumn{2}{c}{\cellcolor{gray!10}de\_de} \\

    \midrule

    \textbf{Female} & \textbf{Time} & \textbf{31} & \textbf{3} & \textbf{9} & \textbf{0} & \textbf{4} & \textbf{6} & \textbf{10} & \textbf{0} & \textbf{0} & \textbf{0} & \textbf{0} & \textbf{0} & \textbf{41} & \textbf{0} & \cellcolor{gray!10}\textbf{1} & \cellcolor{gray!10}\textbf{0} & \cellcolor{gray!10}\textbf{1} & \cellcolor{gray!10}\textbf{0} & \cellcolor{gray!10}\textbf{0} & \cellcolor{gray!10}\textbf{0} & \cellcolor{gray!10}\textbf{2} & \cellcolor{gray!10}\textbf{0} \\
    \midrule
    LibriSpeech I (F) & 24h & 9     & 2     & 8     & -     & 2     & 4     & 4     & -     & -     & -     & -     & -     & 19    & -     & \cellcolor{gray!10}-     & \cellcolor{gray!10}-     & \cellcolor{gray!10}1     & \cellcolor{gray!10}-     & \cellcolor{gray!10}-     & \cellcolor{gray!10}-     & \cellcolor{gray!10}1     & \cellcolor{gray!10}- \\
    LibriSpeech II (F) & 24h & 22    & 1     & 1     & -     & 2     & 2     & 6     & -     & -     & -     & -     & -     & 22    & -     & \cellcolor{gray!10}1     & \cellcolor{gray!10}-     & \cellcolor{gray!10}-     & \cellcolor{gray!10}-     & \cellcolor{gray!10}-     & \cellcolor{gray!10}-     & \cellcolor{gray!10}1     & \cellcolor{gray!10}- \\
    \midrule
    \textbf{Male} & \textbf{Time} & \textbf{33} & \textbf{0} & \textbf{8} & \textbf{0} & \textbf{0} & \textbf{8} & \textbf{8} & \textbf{2} & \textbf{1} & \textbf{0} & \textbf{0} & \textbf{0} & \textbf{46} & \textbf{0} & \cellcolor{gray!10}\textbf{1} & \cellcolor{gray!10}\textbf{0} & \cellcolor{gray!10}\textbf{0} & \cellcolor{gray!10}\textbf{0} & \cellcolor{gray!10}\textbf{0} & \cellcolor{gray!10}\textbf{0} & \cellcolor{gray!10}\textbf{0} & \cellcolor{gray!10}\textbf{0} \\
    \midrule
    LibriSpeech I (M) & 24h & 19    & -     & 3     & -     & -     & 4     & 5     & 2     & -     & -     & -     & -     & 20    & -     & \cellcolor{gray!10}1     & \cellcolor{gray!10}-     & \cellcolor{gray!10}-     & \cellcolor{gray!10}-     & \cellcolor{gray!10}-     & \cellcolor{gray!10}-     & \cellcolor{gray!10}-     & \cellcolor{gray!10}- \\
    LibriSpeech II (M) & 24h & 14    & -     & 5     & -     & -     & 4     & 3     & -     & 1     & -     & -     & -     & 26    & -     & \cellcolor{gray!10}-     & \cellcolor{gray!10}-     & \cellcolor{gray!10}-     & \cellcolor{gray!10}-     & \cellcolor{gray!10}-     & \cellcolor{gray!10}-     & \cellcolor{gray!10}-     & \cellcolor{gray!10}- \\

    \bottomrule
    \end{tabular}}
    \label{tab:gender}
    \begin{flushleft}
    \small \vspace{-0.5em}
    $A$: Accidental triggers; $W$: Wake word said; Gray cells: Mismatch between played audio and wake word model language.\\
    \end{flushleft}
    \vspace{-1.5em}
\end{table*}%

\paragraph{Comparison Across Languages}\label{sec:en-de-zh}
Somewhat expected is the result that the three Chinese and the German smart speaker do not trigger very often on English (US) content (cf. right part of Table~\ref{tab:prevalence}). 
In Table~\ref{tab:language}, we report the results for the differences across languages to explore another potential model bias of the evaluated systems.
Even though we only tested a small number of datasets per language, the number of triggers of VA5~Google and VA6~Apple is very low and comparable to their English performance.
Given the fact that we played the very same episodes of the TV show \emph{Modern Family} in English (US) and German, we find the wake word ``Computer'' to be more resistant to accidental triggers in German ($1$) than in English ($9$). A similar but less pronounced behavior can be seen with ``Alexa.''
Moreover, we found that ``big brother'' in Standard Chinese dàgē (\begin{CJK*}{UTF8}{gbsn}大哥\end{CJK*}) is often confused with the wake word ``Echo'', which is hence not the best wake word choice for this language. Similarly, the German words ``Am Sonntag'' (``On Sunday''), with a high prevalence notably in weather forecasts, are likely to be confused with ``Amazon.''

\subsection{Reproducibility}\label{sec:robustness}
During the verification step of our accidental trigger search, we replayed every trigger $10$~times to measure its reproducibility.  This experiment is designed based on the insight that accidental triggers likely represent samples near the decision thresholds of the machine learning model. Furthermore, we cannot control all potential parameters during the empirical experiments, and thus we want to study if, and to which extent, a trigger is actually repeatable. 

\begin{figure}[!b]
  \centering
  \includegraphics[width=0.85\columnwidth]{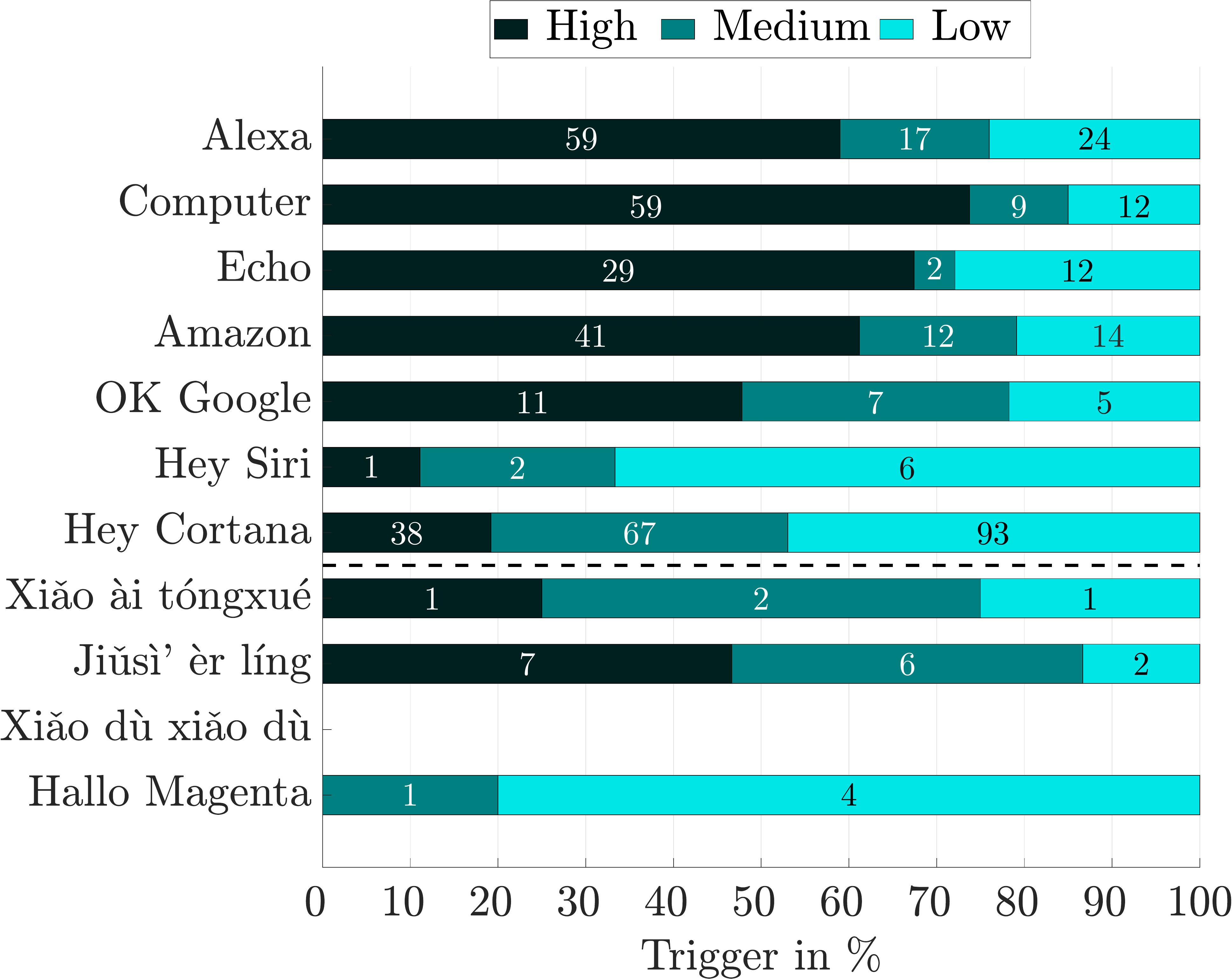}
  \caption{Accidental Trigger Reproducibility. Note that the four speakers below the dashed line do not use wake words in English (US); ``Xiǎo dù xiǎo dù'' did not have any triggers.}
  \label{fig:robustness}
  \vspace{-0.5em}
\end{figure}

We binned the triggers into three categories: \emph{low}, \emph{medium}, and \emph{high}. Audio snippets that triggered the respective assistant 1--3~times are considered as \emph{low}, 4--7~times as \emph{medium}, and 8--10~times as \emph{high}. In Figure~\ref{fig:robustness}, we visualize these results.
We observe that across the Amazon and Google speakers, around 75\,\% of our found triggers are medium to highly reproducible. This indicates that most of the identified triggers are indeed reliable and represent examples where the wake word recognition fails. For the Apple and Microsoft speakers, the triggers are less reliable in our experiments. One caveat of the results is that the Chinese and German speakers' data are rather sparse and do not allow any meaningful observation and interpretation of the results.

%%%%%%%%%%%%%%%%%%%%%%%
%%% LANGUAGE TABLE
%%%%%%%%%%%%%%%%%%%%%%%
\begin{table*}[tbp]
  \centering
  \caption{Differences in Languages.}
  \smallskip
  \resizebox{1.0\textwidth}{!}{
    
\begin{tabular}{rc|cc|cc|cc|cc|cc|cc|cc|cc|cc|cc|cc}
\toprule
\multicolumn{1}{r}{} &   & \multicolumn{2}{c|}{\multirow{2}[2]{*}{\textbf{Alexa}}} & \multicolumn{2}{c|}{\multirow{2}[2]{*}{\textbf{Computer}}} & \multicolumn{2}{c|}{\multirow{2}[2]{*}{\textbf{Echo}}} & \multicolumn{2}{c|}{\multirow{2}[2]{*}{\textbf{Amazon}}} & \multicolumn{2}{c|}{\textbf{Ok}} & \multicolumn{2}{c|}{\textbf{Hey}} & \multicolumn{2}{c|}{\textbf{Hey}} & \multicolumn{2}{c|}{\textbf{Xiǎo ài}} & \multicolumn{2}{c|}{\textbf{Jiǔsì'}} & \multicolumn{2}{c|}{\textbf{Xiǎo dù}} & \multicolumn{2}{c}{\textbf{Hallo}} \\
\multicolumn{1}{r}{} &   & \multicolumn{2}{c|}{} & \multicolumn{2}{c|}{} & \multicolumn{2}{c|}{} & \multicolumn{2}{c|}{} & \multicolumn{2}{c|}{\textbf{Google}} & \multicolumn{2}{c|}{\textbf{Siri}} & \multicolumn{2}{c|}{\textbf{Cortana}} & \multicolumn{2}{c|}{\textbf{tóngxué}} & \multicolumn{2}{c|}{\textbf{èr líng}} & \multicolumn{2}{c|}{\textbf{xiǎo dù}} & \multicolumn{2}{c}{\textbf{Magenta}} \\
\cmidrule{3-24}    \multicolumn{1}{r}{} &   & A     & W     & A     & W     & A     & W     & A     & W     & A     & W     & A     & W     & A     & W     & A     & W     & A     & W     & A     & W     & A     & W \\
\midrule

\textbf{English} & \textbf{Time} & \multicolumn{2}{c|}{\textbf{en\_us}} & \multicolumn{2}{c|}{\textbf{en\_us}} & \multicolumn{2}{c|}{\textbf{en\_us}} & \multicolumn{2}{c|}{\textbf{en\_us}} & \multicolumn{2}{c|}{\textbf{en\_us}} & \multicolumn{2}{c|}{\textbf{en\_us}} & \multicolumn{2}{c||}{\textbf{en\_us}} & \multicolumn{2}{c|}{\cellcolor{gray!10}zh\_cn} & \multicolumn{2}{c|}{\cellcolor{gray!10}zh\_cn} & \multicolumn{2}{c|}{\cellcolor{gray!10}zh\_cn} & \multicolumn{2}{c}{\cellcolor{gray!10}de\_de} \\
\midrule
Modern Family & 24h & 6     & -     & 9     & 4     & 4     & -     & 12    & 1     & 1     & -     & 1     & -     & 23    & \multicolumn{1}{c||}{-}     & \cellcolor{gray!10}-     & \cellcolor{gray!10}-     & \cellcolor{gray!10}1     & \cellcolor{gray!10}-     & \cellcolor{gray!10}-     & \cellcolor{gray!10}-     & \cellcolor{gray!10}1     & \cellcolor{gray!10}- \\
\midrule
\textbf{German} & \textbf{Time} & \multicolumn{2}{c|}{\textbf{de\_de}} & \multicolumn{2}{c|}{\textbf{de\_de}} & \multicolumn{2}{c|}{\textbf{de\_de}} & \multicolumn{2}{c|}{\textbf{de\_de}} & \multicolumn{2}{c|}{\textbf{de\_de}} & \multicolumn{2}{c||}{\textbf{de\_de}} & \multicolumn{2}{c|}{\cellcolor{gray!10}en\_us} & \multicolumn{2}{c|}{\cellcolor{gray!10}zh\_cn} & \multicolumn{2}{c|}{\cellcolor{gray!10}zh\_cn} & \multicolumn{2}{c||}{\cellcolor{gray!10}zh\_cn} & \multicolumn{2}{c}{\textbf{de\_de}} \\
\midrule
Modern Family & 24h & 1    & 1      & 1     & 13   & 3     & -     & 13    & 1     & 2      & -     & 2     & \multicolumn{1}{c||}{-}     & \cellcolor{gray!10}17   & \cellcolor{gray!10}-     &  \cellcolor{gray!10}-    & \cellcolor{gray!10}-      & \cellcolor{gray!10}1     & \cellcolor{gray!10}-     & \cellcolor{gray!10}-     & \multicolumn{1}{c||}{\cellcolor{gray!10}-}     & -     & - \\
\midrule
The Big Bang Theory & 24h & -     & -     & 1     & 9     & 9     & -     & 3     & 2     & 2     & 1     & 1     & \multicolumn{1}{c||}{1}     & \cellcolor{gray!10}12    & \cellcolor{gray!10}-     & \cellcolor{gray!10}-     & \cellcolor{gray!10}-     & \cellcolor{gray!10}-     & \cellcolor{gray!10}-     & \cellcolor{gray!10}-     & \multicolumn{1}{c||}{\cellcolor{gray!10}-}     & 1     & - \\
Polizeiruf 110 & 24h & 3     & -     & 4     & 7     & 3     & -     & 13     & -     & -     & -     & -     & \multicolumn{1}{c||}{-}     & \cellcolor{gray!10}18    & \cellcolor{gray!10}-     & \cellcolor{gray!10}-     & \cellcolor{gray!10}-     & \cellcolor{gray!10}-     & \cellcolor{gray!10}-     & \cellcolor{gray!10}-     & \multicolumn{1}{c||}{\cellcolor{gray!10}-}     & -     & - \\
Tatort & 24h & -     & -     & -     & 8     & 4     & -     & 15     & 1     & 2     & -     & -     & \multicolumn{1}{c||}{-}     & \cellcolor{gray!10}6    & \cellcolor{gray!10}-     & \cellcolor{gray!10}-     & \cellcolor{gray!10}-     & \cellcolor{gray!10}-     & \cellcolor{gray!10}-     & \cellcolor{gray!10}-     & \multicolumn{1}{c||}{\cellcolor{gray!10}-}     & -     & - \\
\midrule
ARD Tagesschau & 12h & 3     & -     & 1     & 1     & -     & -     & 10     & 13     & 1     & -     & -     & \multicolumn{1}{c||}{1}     & \cellcolor{gray!10}29    & \cellcolor{gray!10}-     & \cellcolor{gray!10}1     & \cellcolor{gray!10}-     & \cellcolor{gray!10}-     & \cellcolor{gray!10}-     & \cellcolor{gray!10}-     & \multicolumn{1}{c||}{\cellcolor{gray!10}-}     & -     & - \\
ZDF Heute Journal & 12h & -     & -     & -     & 4     & -     & -     & 5     & 3     & -     & -     & -     & \multicolumn{1}{c||}{-}     & \cellcolor{gray!10}8    & \cellcolor{gray!10}-     & \cellcolor{gray!10}-     & \cellcolor{gray!10}-     & \cellcolor{gray!10}1     & \cellcolor{gray!10}-     & \cellcolor{gray!10}-     & \multicolumn{1}{c||}{\cellcolor{gray!10}-}     & -     & - \\
\midrule
\textbf{Standard Chinese} & \textbf{Time} & \multicolumn{2}{c|}{\cellcolor{gray!10}en\_us} & \multicolumn{2}{c|}{\cellcolor{gray!10}en\_us} & \multicolumn{2}{c|}{\cellcolor{gray!10}en\_us} & \multicolumn{2}{c|}{\cellcolor{gray!10}en\_us} & \multicolumn{2}{c|}{\cellcolor{gray!10}en\_us} & \multicolumn{2}{c|}{\cellcolor{gray!10}en\_us} & \multicolumn{2}{c||}{\cellcolor{gray!10}en\_us} & \multicolumn{2}{c|}{\textbf{zh\_cn}} & \multicolumn{2}{c|}{\textbf{zh\_cn}} & \multicolumn{2}{c||}{\textbf{zh\_cn}} & \multicolumn{2}{c}{\cellcolor{gray!10}de\_de} \\
\midrule
All Is Well & 24h & \cellcolor{gray!10}1     & \cellcolor{gray!10}-     & \cellcolor{gray!10}1     & \cellcolor{gray!10}-     & \cellcolor{gray!10}9     & \cellcolor{gray!10}-     & \cellcolor{gray!10}6     & \cellcolor{gray!10}-     & \cellcolor{gray!10}-     & \cellcolor{gray!10}-     & \cellcolor{gray!10}-     & \cellcolor{gray!10}-     & \cellcolor{gray!10}28    & \multicolumn{1}{c||}{\cellcolor{gray!10}-}     & -     & -     & -     & -     & 2     & \multicolumn{1}{c||}{-}     & \cellcolor{gray!10}-     & \cellcolor{gray!10}- \\
CCTV X.~Lianbo & 12h & \cellcolor{gray!10}3     & \cellcolor{gray!10}-     & \cellcolor{gray!10}1     & \cellcolor{gray!10}-     & \cellcolor{gray!10}1     & \cellcolor{gray!10}-     & \cellcolor{gray!10}-     & \cellcolor{gray!10}-     & \cellcolor{gray!10}-     & \cellcolor{gray!10}-     & \cellcolor{gray!10}1     & \cellcolor{gray!10}-     & \cellcolor{gray!10}38    & \multicolumn{1}{c||}{\cellcolor{gray!10}-}     & -     & -     & 3     & -     & -     & \multicolumn{1}{c||}{-}     & \cellcolor{gray!10}-     & \cellcolor{gray!10}- \\
\bottomrule
\end{tabular}}
    \label{tab:language}
    \begin{flushleft}
    \small \vspace{-0.5em}
    $A$: Accidental triggers; $W$: Wake word said; Gray cells: Mismatch between played audio and wake word model language.\\
    \end{flushleft}
    \vspace{-2em}
\end{table*}%

%% file: crafting.tex
\section{Crafting Accidental Triggers}\label{sec:crafting-triggers}

The previous experiments raise the question of whether it is possible to specifically forge accidental triggers in a systematic and fully automated way. We hypothesize that words with a similar pronunciation as the wake word, i.\,e., based on similar phones (the linguistically smallest unit of sounds) are promising candidates. In this section, we are interested in crafting accidental triggers that are likely caused by the wake word's phonetic similarity.

\subsection{Speech Synthesis}
\label{sec:tts}
To systematically test candidates, we utilize Google's \ac{TTS} API. To provide a variety across different voices and genders, we synthesize $10$ different \ac{TTS} versions, one for each US English voice in the \ac{TTS} API. Four of the voices are standard \ac{TTS} voices; six are Google \emph{WaveNet} voices~\cite{van-den-oord-16-google-wavenet}. In both cases, the female-male-split is half and~half.

Note that some words have more than one possible pronunciation (e.\,g., \mbox{T AH M \textbf{EY} T OW} vs.\ \mbox{T AH M \textbf{AA} T OW}). Unfortunately,  we cannot control how Google's \ac{TTS} service pronounces these words. Nevertheless, we are able to show how, in principle, one can find accidental triggers, and we use $10$ different voices for the synthesis to limit this effect.

\subsection{Levenshtein Distance}

To compare the wake words with other words, we use the Fisher corpus~\cite{cieri-04-fisher} version of the Carnegie Mellon University pronouncing dictionary~\cite{lenzo-14-cmudict}, an open-source pronunciation dictionary for North American English listing the phone sequences of more than $130,000$ words.
We propose two versions of a weighted phone-based Levenshtein distance~\cite{navarro-01-levenshtein} to measure the distance of the phonetic description of a candidate to the phonetic description of the respective wake word in order to find potential triggers in a fully automated way.
Using dynamic programming, we can compute the minimal distance~$\mathcal{L}$ (under an optimal alignment of the wake word and the trigger word). Formally, we calculate
\begin{equation}
\mathcal{L} = \frac{s \cdot S + d \cdot D + i \cdot I}{N}
\label{eq:levensthein}
\end{equation}
with the number of \emph{substituted} phones $S$, \emph{inserted} phones $I$, \emph{deleted} phones $D$, and the total number of phones $N$, describing the weighted \emph{edit distance} to transform one word into another. 
The parameters $s$, $d$, and $i$ describe scale factors for the different kinds of errors. 

In the following, we motivate our different scale factors: During the decoding step of the recognition pipeline, a path search through all possible phone combinations is conducted by the automatic speech recognition system. In general, for the recognition, the path with the least cost is selected as the designated output of the recognition (\ie wake word or not wake word). Considering these principles of wake word recognition, we assume that the different kinds of errors have different impacts on the wake word recognition, as \eg utterances with deletions of relevant phones will hardly act as a wake word.

To find the optimal scale factors, we conducted a \emph{hyperparameter search} where we tested different combinations of weights. For this, we played all different TTS versions of 50,000 English words and measured which of the voice assistants triggered at least once. In total, we were able to measure 826 triggers.
In a second, more advanced, version of this distance measure, we considered phone-depended weights for the different kinds of errors. A more detailed description of this version of the distance measure is presented in Section~\ref{sec:phone_dep}.

We ignore words which are either the wake word itself or pronounced like parts of the wake word (\eg ``Hay'' is blocklisted for ``Hey'' or ``computed'' for ``computer''). The blocklist of the wake words contains a minimum of $2$ words (\emph{Cortana}) and up to $6$ words (\emph{Computer}).
For the optimization, we used a ranked-based assessment: We sorted all $50,000$ words by their distance~$\mathcal{L}$ and used the rank of the triggered word with the largest distance as a metric to compare the different weighted Levenshtein distances. 
With this metric, we performed a grid search for $s$, $d$, $i$ over the interval $[0, 1]$ with a step width of $0.05$.

Note that not all accidental triggers can be explained effectively with the proposed model. Therefore, in a first step, we filter all available triggers to only include those that can be described with this model. This step is necessary as we are not interested in crafting \emph{all} possible accidental triggers such as noise, but accidental triggers that are likely caused by the phonetic similarity to the wake word only.
Also, in case of the Invoke speaker and the Google Home Mini, these speakers have two potential wake words. By focusing on the subset of accidental triggers that can describe the respective wake word more closely, we can filter out the other version of the wake word.
Specifically, we only used triggers were we were able to describe the trigger with the proposed distance measure in such a way that it remained within the first $1$\,\% (500) of words if we overfitted the distance measure to that specific word.  In other words, we only considered triggers for our hyperparameter search where a combination of scale factors exits such that the trigger has at most the rank 500.
After applying this filter criterion, 255 out of the 826 triggers remained in the dataset.

\subsection{Phone-Dependent Weights}
\label{sec:phone_dep}
For a more advanced version of the weighted Levenshtein distance, we utilized information about how costly it is to substitute, delete, and insert specific phones (\ie intuitively it should be less costly to replace one vowel with another vowel in comparison to replacing a vowel with a consonant). 
For this, we calculated phone-dependent weights as described in the following: We used a trained \ac{ASR} system and employed \emph{forced alignment}, which is usually used during the training of an \ac{ASR} system to avoid the need for a detailed alignment of the transcription to the audio file. We can use this algorithm to systematically change phones in the transcription of an audio file and measure the costs of these specific changes. 

\begin{figure}[!tbp]
  \centering
  \includegraphics[width=1.0\columnwidth]{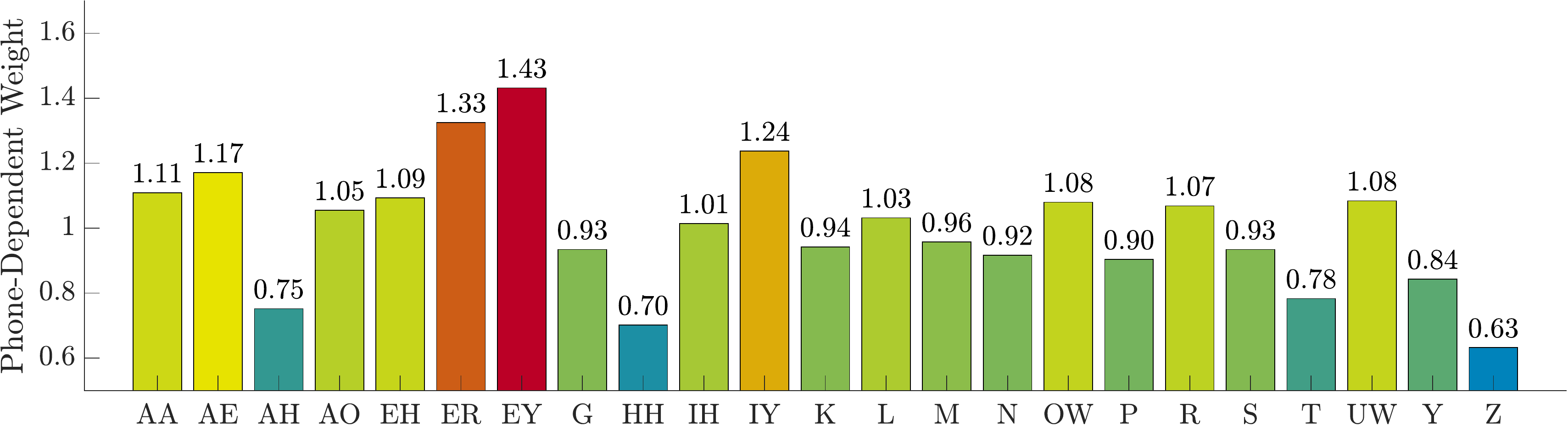}
  \caption{Deletion weights used for the advanced version of the weighted Levenshtein distance. The higher the value, the higher the costs if this phone is \emph{removed}.}
  \label{fig:deletion}
  \vspace{-0.2em}
\end{figure}

To measure the impact of such changes, we distinguish between deletions, substitutions, and insertions: To assess the cost of the deletion of specific phones, we randomly draw $100$ words that contain that specific phone and synthesize $10$ versions of this word via Google's \ac{TTS} API. We use the difference of the scores of the forced alignment output with and without this specific phone for all \ac{TTS} versions of the word. For example, we use the word \emph{little} with the phonetic description L IH T AH L for the phone AH in \emph{Alexa} and measure the score of the forced alignment algorithm for L IH T AH L and L IH T L. The loss in these two scores describes the cost of deleting the sound `AH' in this specific context. For the final weights, we use the average over all $100$ words and $10$ \ac{TTS} versions and finally normalize the values of all averaged phone costs to have a mean value of $1.0$. The deletion weights are shown in Figure~\ref{fig:deletion}.

\begin{figure}[!tbp]
  \centering
  \includegraphics[width=1.0\columnwidth]{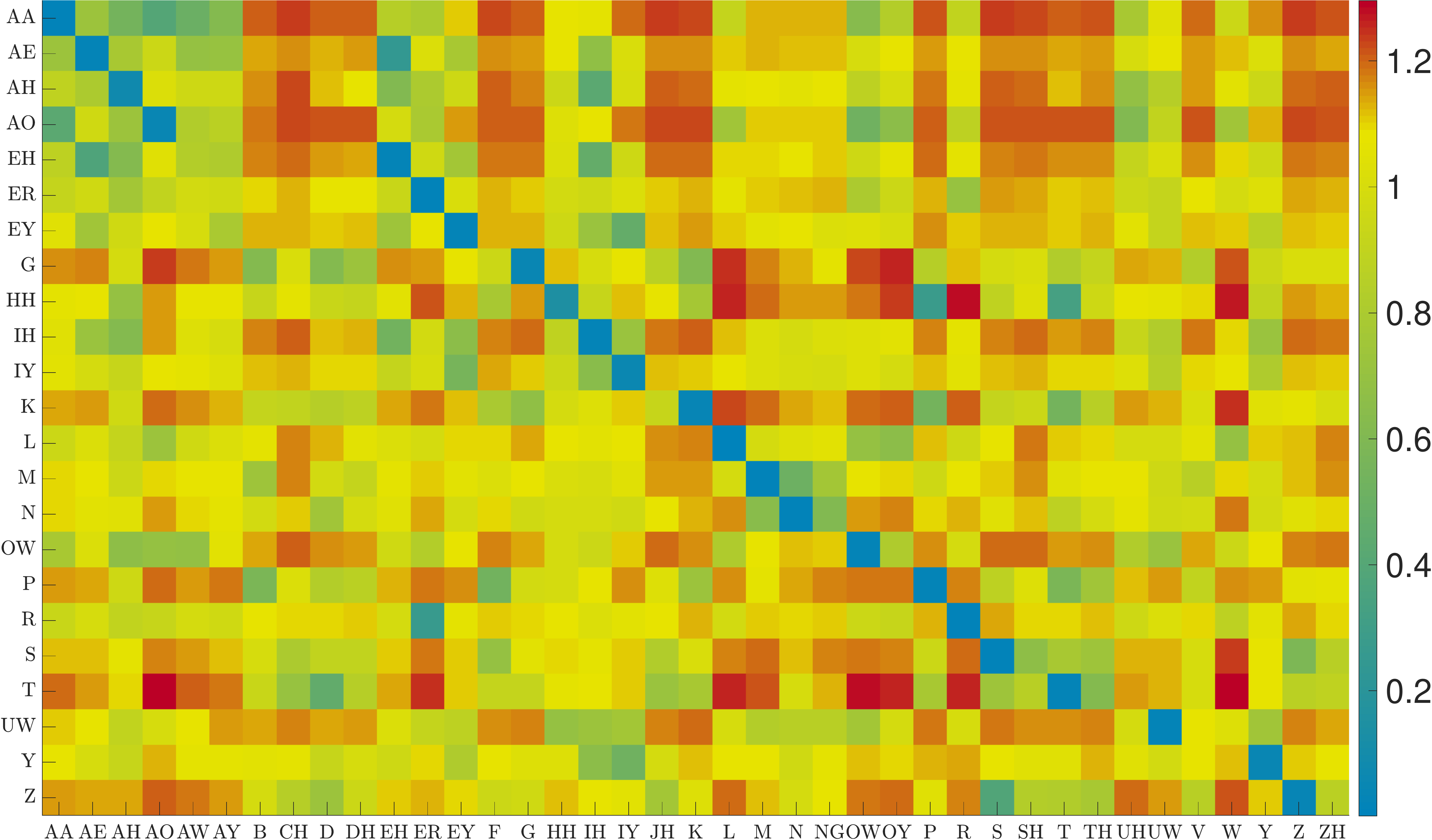}
  \caption{Substitution weights used for the advanced version of the weighted Levenshtein distance plotted as a matrix describing the cost to \emph{replace} the phone in the row with a phone of the columns.}
  \label{fig:substitution}
  \vspace{-0.2em}
\end{figure}

\begin{figure}[!tbp]
  \centering
  \includegraphics[width=1.0\columnwidth]{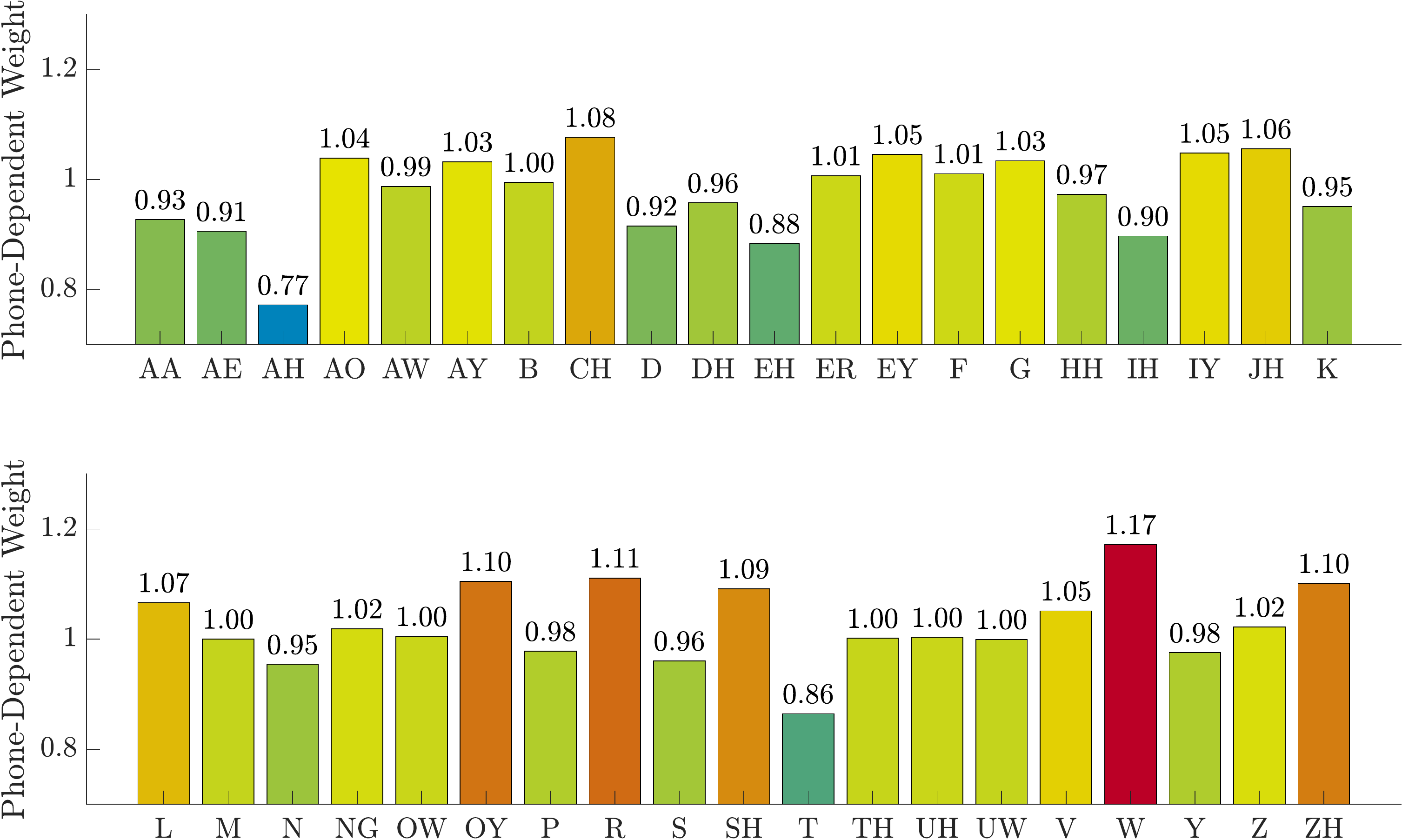}
  \caption{Insertion weights used for the advanced version of the weighted Levenshtein distance. The higher the value, the higher the costs if this phone is \emph{inserted}.}
  \label{fig:insertion}
  \vspace{-0.2em}
\end{figure}

Similarly, to determine the cost of all possible substitutions, we replace the phone-under-test with all other phones for all $100$ words and $10$ \ac{TTS} versions. The matrix of the substitution weights is shown in Figure~\ref{fig:substitution}. Note that we only calculated the weights of phones that occur in the wake words. Therefore, the rows in the figure do not show all theoretically possible
~phones.
The rows of the matrix are also normalized to have an average value of $1.0$.
Finally, we compare the scores between the original transcription and insert the considered phone for the insertion weights. These weights are also normalized to have an average value of $1.0$. The insertion weights are shown in Figure~\ref{fig:insertion}.
All weights are then used along with the scale factors.

\subsection{Cross-Validation}
\label{sec:cross}
We performed a leave-one-out cross-validation to measure the performance of Equation~\eqref{eq:levensthein} in predicting whether words are potential accidental triggers. For this purpose, we compared three different versions of Equation~\eqref{eq:levensthein}: a version, with all scales set to $1$ (Unweighted), a scaled version where we optimized the scale factors (Simple), and a version where we used our optimized scale factors and the phone-dependent weights (Advanced).

\begin{table}[!b]
\caption{Results of the leaving-one-out cross-validation. We report the numbers of triggers within the $100$ words with the smallest distance to the respective wake word.}
\smallskip
\centering
%\resizebox{0.95\columnwidth}{!}{
\begin{tabular}{llc|ccc} 
\toprule 
\textbf{ID} & \textbf{Wake Word} & \textbf{Total} & \textbf{Unweighted} & \textbf{Simple} & \textbf{Advanced} \\  
\midrule 
VA1 & \emph{Alexa}          & 52    & ~9 & 17 & 24 \\
VA2 & \emph{Computer}       & 75    & 17 & 21 & 32 \\
VA3 & \emph{Echo}           & 23    & ~4 & ~5 & 12 \\
VA4 & \emph{Amazon}         & 12    & ~1 & ~1 & ~7 \\
VA5a & \emph{OK Google}     & 2     & ~0 & ~0 & ~0 \\
VA5b & \emph{Hey Google}    & 1     & ~0 & ~0 & ~0 \\
VA6 & \emph{Hey Siri}       & 7     & ~3 & ~3 & ~5 \\
VA7a & \emph{Hey Cortana}   & 38    & ~9 & ~9 & ~6 \\
VA7b & \emph{Cortana}       & 45    & 13 & 14 & 10 \\
\bottomrule
\end{tabular}%}
\label{tab:cross-validation}
\end{table}

We have run a hyperparameter search for the simple and the advanced version of eight wake words triggers for each fold and tested the resulting scale factors on the remaining wake word. The results in Table~\ref{tab:cross-validation} show the number of triggers we find within the $100$ words with the smallest distance for all three versions of the Levenshtein distance and all wake words. Note that the distances tend to cluster words into same distances due to the fixed length of each wake word and, therefore, the same total number of phones $N$, especially for the unweighted and the simple version.

For cases where it is not possible to clearly determine the closest $100$ words, we use all words with a smaller distance than the $100^{th}$ word and draw randomly out of the words with the next largest distance until we obtain a list of $100$ words which makes sure to have a fair comparison in Table~\ref{tab:cross-validation}.

In the third column (Total), we show the total number of words that triggered the perspective wake words, out of the 50,000 words, after filtering. Note that the Google wake words had only $1$ or $2$ triggers and that therefore not more than these can be in the top $100$.

The different versions of the Levenshtein distance generally show better results for the simple and the advanced version compared to the unweighted version, especially for all Amazon wake words. Only for the two wake words from Microsoft, this is not the case. Nevertheless, the advanced version shows the best results on average and is, therefore, the version we use in the following experiments. Notably, for \eg \emph{Computer}, approximately one third (32/100) of the words with the smallest distance actually triggered the smart speaker and for many of the wake words, more or almost half of all possible triggers can be found within the $100$ words with the smallest distance.

\subsection{Performance on Real-World Data}
With the optimized scale factors and weights, we evaluated the distance measure on the transcriptions of the CHiME dataset to assess the performance of the optimized distance measure on real-world voice data. For this purpose, we considered n-grams to test also sequences of words that occur in the CHiME transcriptions, namely 1-, 2-, and 3-grams.

We ran a hyperparameter search for the advanced version of the Levenshtein distance (scale factors and phone-dependent weights) on the triggers of all $9$ wake words on the data set used in Section~\ref{sec:cross}. For these, the optimal scale factors are $s = 1.46$, $d = 1.30$, and $i = 0.24$, which we use for the following experiments.
We select the $100$ words with the smallest distance to the respective wake word from all 1-, 2-, and 3-grams. In total, $300$ n-grams for each wake word. All these n-grams are then synthesized with Google's \acs{TTS} API.
The results of the CHiME n-grams are shown in Table~\ref{tab:chime}. We were able to find a significant number of triggers for almost all of the wake words, like ``fresh parmesan'' for \emph{Amazon}, ``my cereal'' for \emph{Hey Siri}, and ``all acts of'' for \emph{Alexa}.

\begin{table}[!tbp]
\caption{To craft realistic word \emph{combinations}, we construct word sequences based on n-grams from the CHiME transcriptions. We report the numbers of triggers within the $100$ n-grams with the smallest distance to the respective wake word.}
\vspace{0.5em}
\smallskip
\centering
%\resizebox{0.95\columnwidth}{!}{
\begin{tabular}{ll|ccc} 
\toprule 
\textbf{ID} & \textbf{Wake Word} & \textbf{1-gram} & \textbf{2-gram} & \textbf{3-gram} \\  
\midrule 
VA1 & \emph{Alexa}          & ~7 & 10 & 5 \\
VA2 & \emph{Computer}       & 16 & 12 & 10 \\ 
VA3 & \emph{Echo}           & ~1 & ~8 & ~3 \\
VA4 & \emph{Amazon}         & ~2 & 11 & ~4 \\
VA5a & \emph{OK Google}     & ~0 & ~1 & ~0 \\
VA5b & \emph{Hey Google}    & ~0 & ~0 & ~0 \\
VA6 & \emph{Hey Siri}       & ~2 & ~2 & ~0 \\
VA7a & \emph{Hey Cortana}   & ~8 & ~8 & ~4 \\
VA7b & \emph{Cortana}       & ~7 & ~5 & ~6 \\
\bottomrule
\end{tabular}%}
\label{tab:chime}
\vspace{1em}
\end{table}

%% file: related.tex
\section{Related Work}
There is an increasing amount of work focusing on the security and privacy of smart speakers that motivate and guides our research, as discussed in the following.

\subsection{Smart Speaker Privacy}
Malkin et al.~\cite{malkin-19-priv-attitudes} studied the privacy attitudes of 116 smart speaker users. Almost half of their respondents did not know that their voice recordings are stored in the cloud, and only a few had ever deleted any of their recordings.Moreover, they reported that their participants were particularly protective about other people's recordings, such as guests. Besides conversations that include children, financial, sexual, or medical information, \emph{accidentally captured conversations} were named information that should automatically be screened out and not stored. Lau et al.~\cite{lau-18-are-you-listening} studied privacy perceptions and concerns around smart speakers. They found an incomplete understanding of the resulting privacy risks and document problems with incidental smart speaker users. For example, they describe that two of their participants used the audio logs to surveil or monitor incidental users. They noted that current privacy controls are rarely used. For example, they studied why users do not make use of the mute button on the smart speaker. Most of their participants preferred to simply unplug the device and give trust issues and the inability to use the speaker hands-free as reasons not to press the mute button.
Similarly, Abdi \etal~\cite{abdi-19-assistants-perception} explored mental models of where smart speaker data is stored, processed, and shared. Ammari et al. studied how people use their voice assistants and found users being concerned about \emph{random activations} and documented how they deal with them~\cite{ammari-19-alexa-use-fear}.
Huang et al.~\cite{huang-20-shared-speakers} studied users' concerns about shared smart speakers. Their participants expressed worries regarding \emph{voice match false positives}, unauthorized access of personal information, and the misuse of the device by unintended users such as visitors. They confirmed that users perceive external entities, such as smart speaker vendors, collecting voice recordings as a major privacy threat.
Chung et al.~\cite{chung-17-alexa-trust} named \emph{unintentional voice recordings} a significant privacy threat and warned about entities with legitimate voice data access and commercial interests, as well as helpless users not in control of their voice data.
Tabassum et al.~\cite{tabassum-19-user-preference} studied \emph{always-listening} voice assistants that do not require any wake word. Zeng et al.\cite{zeng-17-priv-smart-home} studied security and privacy-related attitudes of people living in smart homes. Their participants mentioned privacy violations and concerns, particularly around audio recordings. In this paper, we study the actual prevalence and implications of accidental triggers with the goal of providing tangible data on this phenomenon, as well as an effective process for assessing trigger accuracy of devices by means of crafting likely accidental triggers.

Recently and concurrently, Dubois et al.~\cite{dubois-20-characterizing-misactivations} published a paper where they played 134 hours of TV shows to measure the prevalence of accidental triggers. Their setup relied on a combination of a webcam, computer vision, and a network traffic-based heuristic. Their work confirms our results in Section~\ref{sec:prevalence}. In contrast to our work, the authors focused only on a comparatively small English TV show dataset and English-speaking smart speakers. They did not consider speakers from other countries, other languages, or other audio datasets. Furthermore, while their work only speculates about regional differences, our reverse engineering of Amazon Echo internals confirms the existence of different wake word models per language, region, and device type (e.\,g., en-US vs.\ en-GB). Finally, we propose a method to craft accidental triggers that enables us to find new triggers systematically.

\subsection{Inaudible and Adversarial Examples}
Adversarial examples against speech recognition systems try to fool the system to output a wrong transcription. For human listeners, the adversarial examples are not at all or only barely distinguishable from benign audio.
In 2016, Carlini et al.~\cite{carlini-16-hidden-voice} have shown that targeted attacks against HMM-only ASR systems are possible. To create their adversarial audio samples, they used an inverse feature extraction. The resulting audio samples were not intelligible by humans. Sch\"onherr et al.~\cite{schoenherr-19-psychoacousitcs} presented an approach where psychoacoustic modeling, which is borrowed from the MP3 compression algorithm, was used to re-shape the perturbations of the adversarial examples. Their approach improves previous work by hiding the changes to the original audio below the human hearing thresholds. Later, the attack was ported to an over-the-air setting by crafting examples that remain robust across different rooms~\cite{schonherr-19-robust}. The accidental triggers identified by our work can be combined with adversarial examples to wake up smart speakers in an inconspicuous way.

%% file: discussion.tex
\section{Discussion}
The results of our experiments suggest possible reasons for the differences across wake words and raise the question of why their vendors have chosen them in the first place. As the underlying problem of accidental triggers, the trade-off between a low false acceptance and false rejection rate is hard to balance and we discuss potential measures that can help to reduce the impact of accidental triggers on the user's privacy.

\subsection{Wake Word}\label{sec:wake-word-discussion}

\paragraph{Properties of Robust Wake Words}
Looking at the number of words in a wake word, one would assume a clear benefit using two words. This observation is supported by the results in Table~\ref{tab:cross-validation}, where ``Cortana'' leads to more triggers than ``Hey Cortana.''
On the contrary, the shortest wake word ``Echo'' has fewer triggers than ``Hey Cortana,'' suggesting that not only the number of words (and phones) itself is important, but the average distance to common words in the respective language. These results suggest that increasing the number of words in a wake word has the same effect as increasing the distance to common words. If we consider the differences in the prevalence of accidental triggers, and that adding an additional word (\eg ``Hey'') comes at close to no cost for the user, we recommend that vendors deploy wake words consisting of two words.

\paragraph{Word Selection}
Amazon shared some details about why they have chosen ``Alexa'' as their wake word~\cite{bort-16-why-alexa-called-alexa}: The development was inspired by the LCARS, the Star Trek computer, which is activated by saying ``Computer.'' Moreover, they wanted a word that people do not ordinarily use in everyday life. In the end, Amazon decided on ``Alexa'' because it sounded unique and used soft vowels and an ``x.'' The co-founder of Apple's voice assistant chose the name ``Siri'' after a co-worker in Norway~\cite{heisler-12-why-siri-called-siri}. Later, when Apple turned Siri from a push-to-talk into a wake word-based voice assistant, the phrase ``Hey Siri'' was chosen because they wanted the wake word to sound as natural as possible~\cite{apple-18-personalized-siri}.
Based on those examples we can see that the wake word choice in practice is not always a rational, technically founded decision, but driven by other factors like marketing as in ``OK Google,'' ``Amazon,'' ``Xiǎo~dù xiǎo~dù,'' or ``Hallo Magenta,'' or based on other motivations such in the case of ``Siri'' or ``Computer.''
Another issue can arise when trying to port a wake word across languages. An example of that is the confusion of dàgē (``big brother'') and ``Echo'' described in Section~\ref{sec:prevalence-results}, and it gets even more complicated in multilingual households~\cite{singh-19-alexa-multilingual}.

\subsection{Countermeasures}

\paragraph{Local On-Device Speech Recognition}
In 2019, Google deployed an on-device speech recognizer on a smartphone that can transcribe spoken audio in real-time without an Internet connection~\cite{he-19-on-device-recognition,kastrenakes-19-pixel-audio-transcription}. We find such an approach to be promising, as it can help to reduce the impact of accidental triggers by limiting the upload of sensitive voice data. After the local ASR detects the wake word, one can imagine a speaker that transcribes the audio and only after being ensured to have detected a user command/question, uploads the short wake word sequence for cloud verification. When both ASR engines agree about the wake word's presence, the command/question is forwarded to the cloud in text or audio form. Coucke et al has described a smart speaker that runs completely offline and is thus private-by-design~\cite{coucke-18-offline-smart-speaker}.

\paragraph{Device-Directed Queries and Visual Cues}
Amazon presented a classifier for distinguishing device-directed queries from background speech in the context of follow-up queries~\cite{mallidi-18-directed-utterance,amazon-20-alexa-follow-up}. While follow-up queries are a convenience feature, one can imagine a similar system that can reduce the number of accidental triggers. Mhaidli et al.~\cite{mhaidli-20-selectively-activate} explored the feasibility to only selectively activate a voice assistant using gaze direction and voice volume level by integrating a depth-camera to recognize a user's head orientation. While this approach constitutes a slight change in how users interact with a smart speaker, it effectively reduces the risk of accidental triggers, by requiring a direct line-of-sight between the user and the device. However, their participants also expressed privacy concerns due to the presence of the camera.

\paragraph{Privacy Mode and Safewords}
Previous work~\cite{lau-18-are-you-listening} has documented the ineffectiveness of current privacy controls, such as the mute button, given the inability to use the speaker hands-free when muted. We imagine a method similar to a \emph{safeword} as a possible workaround for this problem. For this, the speaker implements a \emph{privacy mode} that is activated by a user saying, ``Alexa, please start ignoring me,'' but could, for example, also be activated based on other events such as the time of the day. In the privacy mode, the speaker disables all cloud functionality, including cloud-based wake word verification and question answering.

The speaker's normal operation is then re-enabled by a user saying, ``Alexa, Alexa, Alexa.'' Repeating the wake word multiple times is similar to a behavior observed when parents call their children multiple times, if they do not like to listen, this will feel natural to use. Due to the requirement to speak the somewhat lengthy safeword, accidental triggers will only happen very rarely. We imagine this privacy control to be more usable than a mute button, as the hands-free operation is still possible. As only the wake word is repeated multiple times, we think that vendors can implement this functionality using the local ASR engine.

\paragraph{Increased Transparency}
Another option is to increase transparency and control over the retention periods and individual recordings. In particular, our experience with Microsoft's \emph{Privacy Dashboard} made it clear that vendors need to implement features to better control, sort, filter, and delete voice recordings. Amazon's and Google's web interface already allow a user to filter interactions by date or device easily.
In particular, we imagine a view that shows potential accidental triggers, e.\,g., because the assistant could not detect a question. Currently, accidental triggers are (intentional) not very present, and are easy to miss in the majority of legitimate requests. If accidental triggers are more visible, we hope that users will start to more frequently use privacy controls such as safewords, the mute button, or to request the deletion of the last interaction via a voice command, e.\,g., ``Hey Google, that wasn't for you.'' At first, integrating such a functionality seems unfavorable to vendors, but it can easily be turned into a privacy feature that can be seen as an advantage over competitors.

\subsection{Limitations}

We have neither evaluated nor explored triggers for varying rooms and acoustic environments, \eg distances or volumes. Even if this might influence the reproducibility, this was not part of our study, as we focused on the general number of accidental triggers in a comparable setup across all experiments. This also implies that our results are somewhat tied to the hard- and software version of the evaluated smart speakers.

Our results are subject to change due to model updates for the local ASR or updates of the cloud model. 

Furthermore, we are dealing with a system that is not entirely deterministic, as others already noted~\cite{kumar-18-skillsquatting}. Accidental triggers we mark as local triggers, sometimes overcome the cloud-based recognizer and vice versa.
The findings are mainly based on the English~(US) language; even though we also played a limited set of German and Standard Chinese media, our results are not applicable to other languages or ASR models. 

%% file: conclusion.tex
\section{Conclusion}

In this work, we conduct a comprehensive analysis of accidental triggers in voice assistants and explore their impact on the user's privacy. We explain how current smart speakers try to limit the impact of accidental triggers using cloud-based verification systems and analyze how these systems affect users' privacy. More specifically, we automate the process of finding accidental triggers and measure their prevalence across $11$~smart speakers. We describe a method to artificially craft such triggers using a pronouncing dictionary and a weighted phone-based Levenshtein distance metric that can be used to benchmark smart speakers. As the underlying problem of accidental triggers, the trade-off between a low false acceptance and false rejection rate is hard to balance. We discuss countermeasures that can help to reduce the number and impact of accidental triggers. To foster future research on this topic, we publish a data set of more than 1000 accidental triggers.

\AddToShipoutPicture*{
     \AtTextUpperLeft{
         \put(380,-710){\emph{``Alexa, Stop!''}}
     }
}
\clearpage